  \pdfoutput=1
  \documentclass[graphicx,amssymb,amsmath,enumerate,11pt]{report}
  \usepackage{fancyhdr}
  \fancyheadoffset[]{0.1cm}

  \usepackage{epsfig}
  \usepackage{url}
  \newcommand\mchapter[2]{\chapter*{#1}
  \vskip -0.5cm \noindent {\it \LARGE #2}
  \addcontentsline{toc}{chapter}{#1\\{\normalsize\it #2}}}

\def\slashchar#1{\setbox0=\hbox{$#1$}
   \dimen0=\wd0 \setbox1=\hbox{/} \dimen1=\wd1
   \ifdim\dimen0>\dimen1 \rlap{\hbox to \dimen0{\hfil/\hfil}} #1
   \else  \rlap{\hbox to \dimen1{\hfil$#1$\hfil}} / \fi}

\def\ereco{E_{\rm rec}}

\def\Journal#1#2#3#4{{#1} {\bf #2}, #3 (#4)}

\def\PLB{{\rm Phys. Lett.}  B}

\def\be{\begin{equation}}
\def\ee{\end{equation}}
\def\bea{\begin{eqnarray}}
\def\eea{\end{eqnarray}}

\newcommand{\bd}{\begin{displaymath} }
\newcommand{\ed}{\end{displaymath} }

\newcommand{\nub}{\overline{\nu}}

\newcommand{\ubar}{\overline{u}}
\newcommand{\dbar}{\overline{d}}
\newcommand{\cbar}{\overline{c}}
\newcommand{\sbar}{\overline{s}}

\newcommand{\qsq}{\mbox{$Q^2$}}

\newcommand{\AmS}{{\protect\the\textfont2
  A\kern-.1667em\lower.5ex\hbox{M}\kern-.125emS}}

\def\lsim{\mathrel{\hbox{\rlap{\hbox{\lower4pt\hbox{$\sim$}}}\hbox{$<$}}}}
\def\gsim{\mathrel{\rlap{\lower4pt\hbox{\hskip1pt$\sim$}} \raise1pt\hbox{$>$}}}

\def\eqref#1{Eq.~\ref{#1}}

\begin{document}      

 \rhead{\bfseries Neutrino-Nucleus Interactions}

\mchapter{Recent Developments in Neutrino/Antineutrino - Nucleus Interactions}
 {Jorge G.~Morf\'{\i}n$^a$, Juan  Nieves$^b$, Jan T. Sobczyk$^a$
}
 \label{ch-23:interactions}

\vspace{0.5cm}

\begin{center}
$^a$ {\it Fermi National Accelerator Laboratory, Batavia, Illinois 60510, USA.  } \\ [6pt]
$^b$ {\it Instituto de Fisica Corpuscular (IFIC), Centro Mixto Universidad de Valencia-CSIC,\\
Institutos de Investigacion de Paterna, E-46071 Valencia, Spain.
}
\end{center}

\vspace{3cm}
\begin{center}
{\bf Abstract}
\end{center}
Recent experimental results and developments in the theoretical treatment of neutrino-nucleus interactions 
in the energy range of 1-10 GeV are discussed. Difficulties in extracting neutrino-nucleon cross sections 
from neutrino-nucleus scattering data are explained and significance of understanding nuclear effects 
for neutrino oscillation experiments is stressed. Detailed discussions of the status of two-body current 
contribution in the kinematic region dominated by quasi-elastic scattering and specific features 
of partonic nuclear effects in weak DIS scattering are presented.

\vskip -18cm
\hfill FERMILAB-PUB-12-529-PPD
\newpage

\section{Introduction}
\label{sec23: intro}
Recent interest in neutrino interactions in the few GeV energy region comes from neutrino oscillation
experiments and their need to reduce systematic errors. 
Neutrino fluxes used in contemporary long and short baseline experiments (K2K, T2K, MINOS, NOvA, MiniBooNE) 
are peaked in the 1 - 5 GeV energy 
domain and during the last $\sim 10$ years there has been considerable theoretical and experimental activity in
the investigation of neutrino cross sections in this domain with reference~\cite{Gallagher:2011zz} being a good summary of the lower-energy situation. 
Several new cross section measurements have been performed by neutrino oscillation collaborations and there are two dedicated cross section experiments (SciBooNE and MINERvA) which have been launched at Fermilab.

Even with this degree of activity, the precision with which the basic neutrino-{\em nucleon} cross sections are known 
is still not better than $20-30$\%. There are two main reasons for this: the poor knowledge of neutrino
fluxes and the fact that all the recent cross section measurements have been performed on nuclear targets.  It is important to recall that what current neutrino experiments are measuring are events that are a convolution of energy-dependent neutrino flux $\otimes$  energy-dependent cross section $\otimes$ energy-dependent nuclear effects.  The experiments have, for example, then measured an  effective neutrino-carbon cross sections and to extract a neutrino-nucleon cross sections from these measurements requires separation of nuclear physics effects that can be done with only limited precision. For many oscillation experiments, using the same nuclear targets for their near and far detectors is a good start.  However, even with the same nuclear target near-and-far, that there are different near and far neutrino energy spectra due to oscillations means there is a different convolution of cross section $\otimes$ nuclear effects near and far and there is no automatic cancellation between the near-and-far detectors.
For a thorough comparison of measured neutrino-nucleon cross sections with theoretical models, these convoluted effects have to be understood.  

Some of the new cross section measurements raised doubts in the areas which seemed to be well understood.  The list of new {\it puzzles} is quite long
and seems to be expanding. What is the value of the quasielastic axial mass? How large is the two-body current contribution that can mimic genuine quasielastic interactions? How large is CC (charged current) coherent pion production at a few GeV neutrino energies? What is behind the large discrepancy between MiniBooNE pion production measurements and theoretical model predictions? It can be seen as a paradox that the more than 30-year old ANL and BNL low statistics deuterium pion production
data, with its minimal nuclear corrections, is still used as the best source of information about the nucleon-$\Delta$ transition matrix element.

Analysis of neutrino scattering data is certainly more complicated than the analysis of electron scattering data.  In the electron case one knows exactly the initial electron energy and so also the values of 
energy- and momentum-transfer.  It is then possible to explicitly study separate interesting kinematical regions like QE (quasielastic) peak or the $\Delta$ peak. Neutrino scattering data is always flux (often wide band!) integrated. Interacting neutrino energy must be evaluated based on kinematics of particles in the final state taking into account detector acceptance and measurement accuracy.

For neutrino-{\it nucleon} interactions one can distinguish: Charged Current Quasielastic (CCQE), Neutral Current elastic (NCEl), Resonance production (RES) and more inelastic reactions up to the deep-inelastic (a rather misleading "DIS"
term is often used to describe all the interactions which are neither CCQE/NCEl nor RES) domain. 
Quite different theoretical tools are used to model each of them. The simplest neutrino hadronic reaction is the charge current quasielastic (CCQE) interaction: $ \nu_\ell + n \rightarrow \ell^- + p$ with two particles: charged lepton and proton in the final state. One would like to extend this definition to the neutrino-nucleus interaction occurring on bound neutrons. The obvious question arises: what is the experimental signature of CCQE on a nuclear target? The ejected proton is not necessarily seen in a detector because quite often its momentum is below the acceptance threshold. However, events with a single reconstructed  charged lepton track can result from a variety of initial interactions eg. from a two body current interaction or from real pion production and its subsequent absorption. Similar problems arise in other type of interactions. It is becoming clear that interpretation of
neutrino-nucleus interaction must rely on a careful data/Monte Carlo (MC) comparison done with reliable MC neutrino event generators. This is why we decided to include in the review some information about development of MC event generators. 

From the experimental point of view it is natural to speak about events with no pions in the final state, with only one pion etc. In fact, in several recent experimental measurements that investigated quantities  defined in this way, the dependence on assumptions of Monte Carlo event generators were minimal.  To compare with the experimental data given in this format one must add contributions from various dynamical mechanisms and also to model FSI effects. Several ingredients of the theoretical models are verified simultaneously. It is clear that in order to validate a model one needs many samples of precise neutrino-nucleus scattering measurements on variety of nuclear targets with various neutrino fluxes. 

Our review is organized as follows, we review recent inclusive measurements in the lower E region and then concentrate on exclusive states in increasing W, the mass of the hadronic system.  Due to the limited length of this review, we do have to limit our coverage to only the most recent developments.
\section{Neutrino Charged Current and Neutral Current Inclusive Reactions} 
\label{sec23:neu}
\subsection{Recent measurements}
\begin{figure}[tbh]
\centerline{\includegraphics[height=8.0cm]{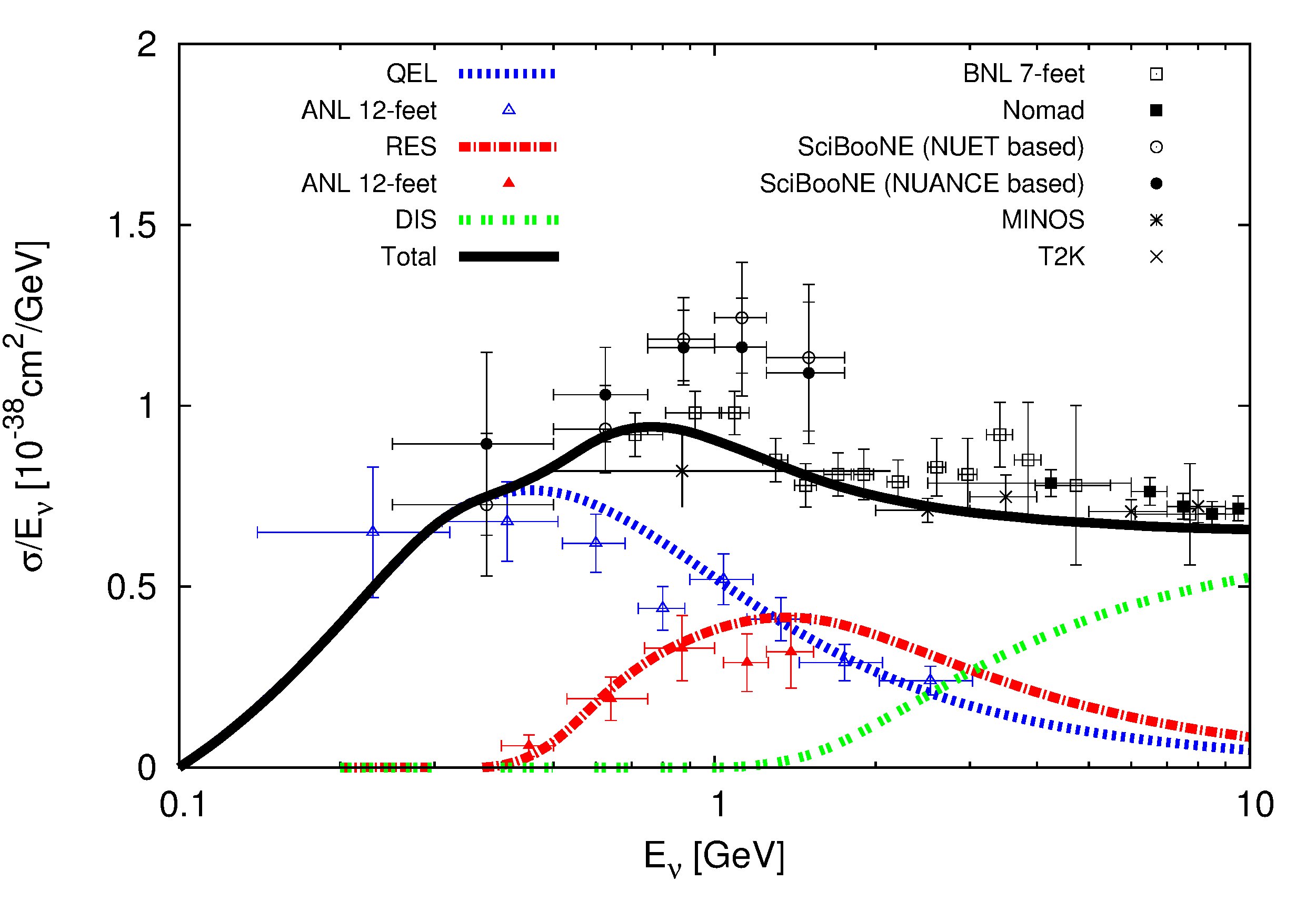}}
\caption{\footnotesize Breakdown of the inclusive CC muon neutrino cross section on free isoscalar target to QE, RES and DIS contributions, as viewed by NuWro MC event generator. }\label {fig23:inclusive}
\end{figure}

There are four recent CC inclusive neutrino and antineutrino cross sections measurements in the $E_\nu \leq 10$~GeV energy region  \cite{23-inclusive}, see Fig. \ref{fig23:inclusive}.  We notice a mild tension between SciBooNE and T2K measurements.  In the following sections QE, RES and DIS contributions will be discussed separately. 
\subsection{Theory. General formulae: outgoing lepton differential cross sections }
In this paper, we will discuss the neutrino CC or NC (neutral current) inclusive reaction:
\begin{equation}
\nu_\ell (k) +\, A_Z \to \ell (k^\prime) + X   .
\label{eq23:reac}
\end{equation}
The generalization of the expressions to antineutrino induced reactions  is straightforward. In the equation above, the outgoing lepton could be either a negatively charged lepton, $\ell^-$, of flavor $\ell$ or a neutrino $\nu_\ell$, for CC or NC processes, respectively.

The double differential cross section, with respect to the outgoing lepton kinematical variables,  for the process of Eq.~(\ref{eq23:reac}) is given in the Laboratory (LAB) frame by
\begin{equation}
\frac{d^2\sigma_{\nu_\ell \ell}}{d\Omega(\hat{k^\prime})dE^\prime_\ell} =
\frac{|\vec{k}^\prime|}{|\vec{k}~|}\frac{G^2_F}{4\pi^2 \eta} 
L^{(\nu)}_{\mu\sigma}W^{\mu\sigma}  \label{eq23:sec}
\end{equation}
with $\vec{k}$ and $\vec{k}^\prime~$ the LAB lepton momenta, 
$E^{\prime}_\ell = (\vec{k}^{\prime\, 2} + m_\ell^2 )^{1/2}$ and $m_\ell$ 
the energy and the mass of the outgoing lepton, $G_F=1.1664\times 10^{-11}$ MeV$^{-2}$, 
the Fermi constant and $L$ and $W$ the leptonic and hadronic tensors, respectively. 
Besides, $\eta$ takes the values 1 or 4 for CC or NC processes, respectively. The leptonic tensor is given by (in this convention,  $\epsilon_{0123}= +1$ and the metric is $g^{\mu\nu}=(+,-,-,-)$):
\begin{eqnarray}
L^{(\nu)}_{\mu\sigma}&=& L^s_{\mu\sigma}+ {\rm i} L^a_{\mu\sigma} =
 k^\prime_\mu k_\sigma +k^\prime_\sigma k_\mu
- g_{\mu\sigma} k\cdot k^\prime + {\rm i}
\epsilon_{\mu\sigma\alpha\beta}k^{\prime\alpha}k^\beta \label{eq23:lep}
\end{eqnarray}
The hadronic tensor includes a collection of non-leptonic vertices and corresponds to the charged or neutral electroweak transitions of the target nucleon or nucleus, $i$, to all possible final states.   It is thus given by
\begin{eqnarray}
W^{\mu\sigma} &=& \frac{1}{2M_i}\overline{\sum_f } (2\pi)^3
\delta^4(P^\prime_f-P-q) \langle f | j^\mu_{\rm cc+,\, nc}(0) | i \rangle
 \langle f | j^\sigma_{\rm cc+,\, nc}(0) | i \rangle^*
\label{eq23:wmunu}
\end{eqnarray}
with $P^\mu$ the four-momentum of the initial target, $M_i^2=P^2$  the target mass square, $P_f^\prime$  
the total four momentum of the hadronic state $f$ and $q=k-k^\prime$ the four momentum transferred to the hadronic 
system.  The bar over the sum denotes the average over initial spins.

The hadronic tensor is completely determined by six independent, Lorentz scalar and real, structure functions $W_i(q^2, q\cdot P)$,
\begin{equation}
\frac{W^{\mu\nu}}{2M_i} = - g^{\mu\nu}W_1 + \frac{P^\mu
  P^\nu}{M_i^2} W_2 + {\rm i}
  \frac{\epsilon^{\mu\nu\gamma\delta}P_\gamma q_\delta}{2M_i^2}W_3 +  
\frac{q^\mu  q^\nu}{M_i^2} W_4 + \frac{P^\mu q^\nu + P^\nu q^\mu}
{2M_i^2} W_5+ {\rm i}\frac{P^\mu q^\nu - P^\nu q^\mu}
{2M_i^2} W_6
\end{equation}
Taking $\vec{q}$ in the $z$ direction and $P^\mu = (M_i, \vec{0})$, it is straightforward to  find the six structure functions in terms of the $W^{00}, W^{xx}=W^{yy}, W^{zz}, W^{xy}$ and $W^{0z}$ components of the hadronic tensor.  After contracting with the leptonic tensor, one obtains that for massless leptons only three of them are relevant, namely
\begin{eqnarray}
\frac{d^2\sigma_{\nu_\ell \ell}}{dx dy} &=&
\frac{G_F^2 M E_\nu}{\eta \pi} \left \{ \left(1-y-\frac{Mxy}{2E_\nu}\right)F_2^{\nu} 
+ xy^2 F_1^{\nu} + y (1-y/2) x F_3^{\nu} 
\right\} \label{eq23:cross}
\end{eqnarray}
with $E_\nu$ the incoming neutrino energy, $M$ the nucleon mass, $x=-q^2/2Mq^0$, $y= q^0/E_\nu$, 
while the nuclear structure functions $F_{1,2,3}^{\nu}$ are given by, $F_1^{\nu}=2 M M_i W_1$,  
$F_2^{\nu}=2(q\cdot P)  W_2$ and $F_3^{\nu}/M=-2(q\cdot P)  W_3/M_i$. The cross section for the CC 
antineutrino induced nuclear reaction is easily obtained by i) changing the sign of the parity-violating term, 
proportional to $F_3$, in the differential cross section (this is because 
$L_{\mu\sigma}^{(\bar \nu)}=L_{\sigma\mu}^{(\nu)}$.), Eq.~(\ref{eq23:cross}), and 
ii) using $j^\mu_{\rm cc-}=j^{\mu\dagger}_{\rm cc+}$ in the definition/computation of the hadron tensor in 
Eq.~(\ref{eq23:wmunu}).  In the case of antineutrino NC driven processes, it is only needed to flip the sign of 
the term proportional to $F_3$ in the differential cross section, since the hadron NC is not affected. 

The hadronic tensor is determined by the $W$ or $Z$ gauge boson selfenergy, $\Pi^{\mu\rho}_{W,Z}(q)$, in the nuclear medium. To evaluate this latter object requires a theoretical scheme, where the relevant degrees of freedom and nuclear effects could be taken into account. 

In the next two sections we will discuss CCQE and pion production reaction. The general formalism described above will be used in the section devoted to DIS.
\section{Charged Current Quasielastic}
As discussed in the Introduction, we define CCQE as either the reaction on a free nucleon or on a 
quasi free nucleon inside a nucleus yielding a muon and nucleon. In the case of neutrino nucleus scattering we also use the term 
CCQE-like reaction defined as one in which there are no pions in the final state.  
It then includes events with real pion production followed by absorption. Such a definition may seem 
awkward but as will be seen, it is close to what was experimentally measured by the MiniBooNE collaboration.         

A theoretical description of the free nucleon target CCQE reaction  is based on the conserved vector current (CVC) and the partially conserved axial current (PCAC) hypotheses. The only unknown quantity is the nucleon axial form-factor $G_A(Q^2)$ for which one typically assumes a dipole form $G_A(0)(1+\frac{Q^2}{M_A^2})^{-2}$ with one free parameter, the axial mass $M_A$. The non-dipole axial form factor was investigated e.g. in \cite{23-nondipole}.

In the past, several measurements of $M_A$ were performed on a deuterium target for which 
most of nuclear physics complications are minimal and it seemed that the results converged to a value of the 
order of $1.03$~GeV \cite{23-bodek_MA}. There is an additional  argument in favor of a similar value of $M_A$ 
coming from the weak pion-production at low $Q^2$. PCAC based evaluation gives an axial mass value of  
$1.077\pm 0.039$~GeV~\cite{23-Bernard:2001rs}. On the other hand, all of the more recent high statistics 
measurements of  $M_A$, with the exception of the NOMAD higher-energy experiment, reported larger values: 
K2K (oxygen, $Q^2>0.2$~GeV$^2$) $\rightarrow 1.2\pm 0.12$~\cite{23-k2k_oxygen_MA}; K2K (carbon, 
$Q^2>0.2$~GeV$^2$) $\rightarrow 1.14\pm 0.11$~\cite{23-k2k_carbon_MA}; MINOS (iron, $Q^2>0$~GeV$^2$) 
$\rightarrow 1.19\pm 0.17$; MINOS (iron, $Q^2>0.3$~GeV$^2$) $\rightarrow 1.26\pm 0.17$~\cite{23-minos_MA}; 
MiniBooNE  (carbon, $Q^2>0$~GeV$^2$) $\rightarrow 1.35\pm 0.17$~\cite{23-AguilarArevalo:2010zc}; 
MiniBooNE (carbon,  $Q^2>0.25$~GeV$^2$) 
$\rightarrow 1.27\pm 0.14$ (for completness:  NOMAD (carbon, $Q^2>0$~GeV$^2$)
$\rightarrow 1.07\pm 0.07$~\cite{23-Lyubushkin:2008pe}).

The difference between MiniBooNE and NOMAD measurements could come from different definitions of the CCQE signal.   
In the case of MiniBooNE a sample of 2-subevents (Cherenkov light  from muon and from decay electron) 
is analyzed and ejected protons  are not detected. In the case of NOMAD 1-track (muon) and 2-tracks (muon and proton) 
samples of events are analyzed simulateuosly. With a suitable chosen value of the formation zone parameter $\tau_0$ 
values of  $M_A$ extracted separately from both data samples are approximately  the same, 
see Table 9 in \cite{23-Lyubushkin:2008pe}.
We note that the procedures in which the formation zone concept is applied to nucleons that already exist may seem 
little controversial. 
We would like to mention also the CCQE data not yet published in peer review journals. 
MINOS tried to evaluate better the pion production background~\cite{23-minos_nuint11}. 
A function of $Q^2$ which  corrects Monte Carlo (NEUGEN) RES predictions was proposed. 
The shape of the curve is similar to MiniBooNE's DATA/MC correction function (see below) 
but in the case of MiniBooNE 
for $Q^2>0.1$~GeV$^2$  the correction factor is $>1$. The new MINOS best fit value of $M_A$ is $1.16$~GeV and 
the error was reduced by a factor of $3$ with respect
to \cite{23-minos_MA}.
SciBooNE showed  partial results of the CCQE analysis~\cite{23-sciboone_nuint11}. 
Results are given in terms of fits for CCQE cross-section DATA/MC multiplicative factors 
$a_j$ ($j$ label true neutrino energy bins) and a scaling factor $F_N$.  The obtained best fit values in 
the neutrino energy region $E_\nu\in (0.6, 1.6)$~GeV  are between $1.00$ and $1.09$ which with $F_N=1.02$ 
and the value of the axial mass used in the NEUT Monte Carlo generator ($1.2$~GeV)  
should translate to the axial mass value $M_A\sim 1.25 - 1.3$~GeV. In the SciBooNE analysis  
there are some instabilities in the wider region of $E_\nu$ (see Fig. 11.2 in \cite{23-alcaraz_thesis}). 
A use of the universal background scaling factor $a_{bcg}$ for three different event samples 
is perhaps not sufficient (its best fit value is as large as $1.37$ GeV). 

An important antineutrino CCQE measurement was reported by MiniBooNE \cite{23-miniboone_nuint11}. The DATA/MC average  cross-section ratio was reported to be $1.21\pm 0.12$ which is a surprising result because in the NUANCE carbon CCQE computations the $M_A$ value was set to be $1.35$~GeV.  In the experimental analysis, it was important to evaluate correctly neutrino contamination in the anti-neutrino flux. Three independent measurements indicate that the $\nu_\mu$ flux in the antineutrino beam should be scaled down by a factor of $\sim 0.8$ with an obvious important impact on the final results.

The most recent MINERvA preliminary results for CCQE antineutrino reaction are still subject to large flux normalization uncertainties but they seem to be consistent with $M_A=0.99$~GeV \cite{23-minerva_ccqe}.
\subsection{MiniBooNE data}
In recent discussions of the CCQE, MiniBooNE measurement plays a special role. 
For the first time the data was presented in the form of double differential cross section 
in muon scattering angle and kinetic energy. Such data is the actual observable for the 
MiniBooNE experiment and more complete than a distribution of events in $Q^2$ which is 
calculated assuming an obviously incorrect nuclear model (the nucleon is assumed to be at rest).  
The signal events form a subset of events with no pions in the final state. 
MiniBooNE subtracted as a background, events with real pion production and 
subsequent absorption and also a contribution from pionless $\Delta$ decays implemented in the 
NUANCE MC~\cite{23-nuance} as constant fractions of $\Delta ^{++}$ and $\Delta^+$ decays, following the approach of 
Ref.~\cite{23-Oset:1987re}. The background estimate, based on MC predictions, was later corrected by a
$Q^2$ dependent function, which accounts for a data/MC discrepancy in the sample of events 
containing one $\pi^+$ in the final state. The shape of the correction function is not well understood 
\cite{23-jarek} but it has an important impact on the extracted value of $M_A$. The function quantifies 
a lack of understanding of processes like pion absorption and can have a significant effect on 
the understanding of both samples of events.  

MiniBooNE also provided data for the CCQE signal plus background together as the 
measurement of the cross section of the process in which there are no pions in the final state, 
the observable which is  maximally independent of MC assumptions.  
\subsection{Theoretical approaches to CCQE - generalities}
\begin{figure}[tbh]
\centerline{\includegraphics[height=8.0cm]{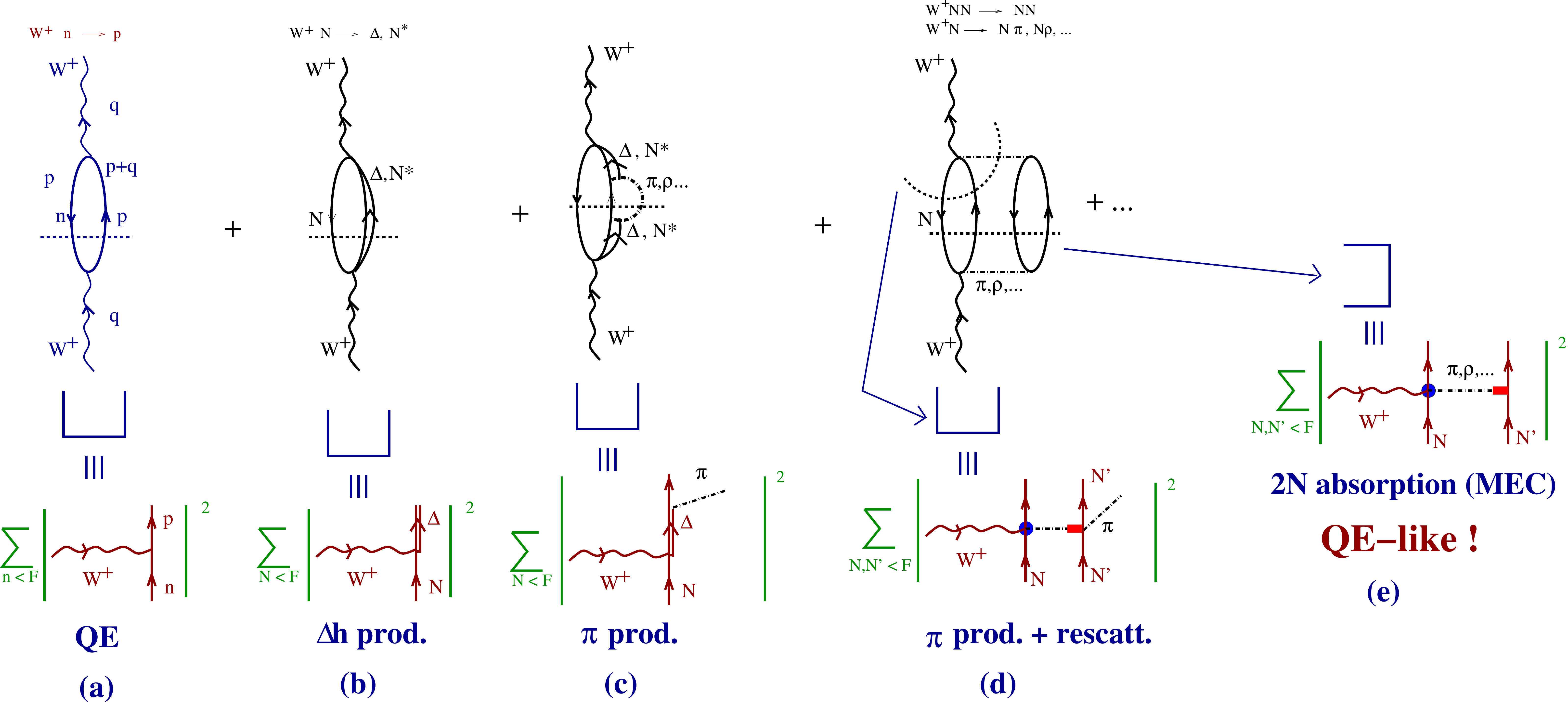}}
\caption{\footnotesize Diagrammatic representation of some diagrams
  contributing to the $W^+-$selfenergy. }\label{fig-23:fig2}
\end{figure}

Several approaches have been followed/derived to compute the relevant
gauge boson absorption modes (self-energy) to describe the CCQE process.  For
moderate and intermediate neutrino energies, in the few GeV region, the most relevant ones are: the absorption
by one nucleon, or a pair of nucleons or even three nucleon
mechanisms, real and virtual meson ($\pi$, $\rho$, $\cdots$)
production, excitation of $\Delta$ of higher resonance degrees of
freedom, etc. (for example, some absorption modes are depicted in
Fig.~\ref{fig-23:fig2} for the case of neutrino CC processes).  
A review of theoretical model results can be found in~\cite{23-Boyd:2009zz}. Almost all approaches, used
at intermediate neutrino
energies, deal with hadron, instead of quarks and gluons, degrees of
freedom.  In addition they consider several nuclear effects
such as RPA or Short Range Correlations (SRC).
The free space couplings between hadrons and/or the weak $W$ and $Z$ bosons are parametrized in terms of
form factors, which are fitted to the available data on electroweak
scattering off free nucleons. In the few GeV energy region,  theoretical models rely on the impulse approximation (IA)
and neutrino-nucleus CCQE interactions are viewed as
a two step process: primary interaction and Final State Interactions (FSI),
propagation 
of resulting hadrons through the nucleus. The validity of the IA is usually related to
typical values of the momentum transfer $q$. Experience from the electron scattering tells us 
that for $q>300-500$~MeV/c IA based models are able to reproduce the data well. Thus, the expectations is that for a few GeV neutrino interactions 
IA is an acceptable approach and if necessary simpler nuclear models computations can be supplemented
with RPA corrections for lower momentun transfers (see below). In the neutrino nucleus cross section measurements a goal is to 
learn about neutrino free nucleon target scattering parameters (an obvious exception is coherent pion
production). {\it Effective} parameters like 
sometimes discussed quasi elastic axial mass $M_A^{eff}$ are of little use as their values can depend on the neutrino
flux, target and perhaps also on the detection technique/acceptance. 

The definition of neutrino-nucleus  CCQE scattering can be made more rigorous in the language of
many body field theory. CCQE process originates from a first step
mechanism where the gauge boson is being absorbed by just one
nucleon. This corresponds to the first of the selfenergy diagrams
depicted in Fig.~\ref{fig-23:fig2} (contribution (a)). This contribution,
that from now on we will call {\it genuine} QE,  has been computed
within different theoretical models and used to predict the
corresponding outgoing
lepton differential cross section. 

The simplest model, commonly used in Monte Carlo event generators, is the relativistic 
Fermi gas (RFG) model  proposed by Smith and
Moniz more than 35 years ago~\cite{23-Smith:1972xh} corresponding to only one many body
Feynman diagram. The model combines the bare nucleon physics
with a model to account for Fermi motion and nucleon binding within the specific
nucleus. 
The model can be made more realistic in many ways\footnote{When the axial mass and electromagnetic form factors are
kept unchanged, the inclusion of more sophisticated nuclear effects
makes the cross section generally smaller with respect to the RFG (relativistic Fermi gas) 
model. } to achieve
better agreement with a broad range of electron scattering data. For
example, the inclusion of a realistic joint distribution of target
nucleon momenta and binding energies based on short range correlation
effects leads to the spectral function (SF) approach. 
Spectral functions for
  nuclei, ranging from carbon (A = 12) to iron (A = 56)
  have been modeled using the Local Density Approximation
  (LDA)~\cite{23-Benhar:1994hw}, in which the experimental information
  obtained from nucleon knock-out measurements is combined with the
  results of theoretical calculations in nuclear matter at
  different densities, and they have been extensively validated with
  electron scattering data. Calculations by Benhar et
  al.,~\cite{23-Benhar:2006nr} and Ankowski et
  al..~\cite{23-Ankowski:2007uy} show that the SF effects moderately
  modify the muon neutrino differential cross sections, and they lead
  to reductions of the order of 15\% in the total cross sections. This is
  corroborated by the results obtained within the semi-phenomenological
  model (density dependent mean-field potential in which the nucleons
  are bound)~\cite{23-Leitner:2008ue} employed within the GiBUU model to
  account for these effects.  

\begin{figure}[htb]
\begin{center}
\makebox[0pt]{\includegraphics[height=6.0cm]{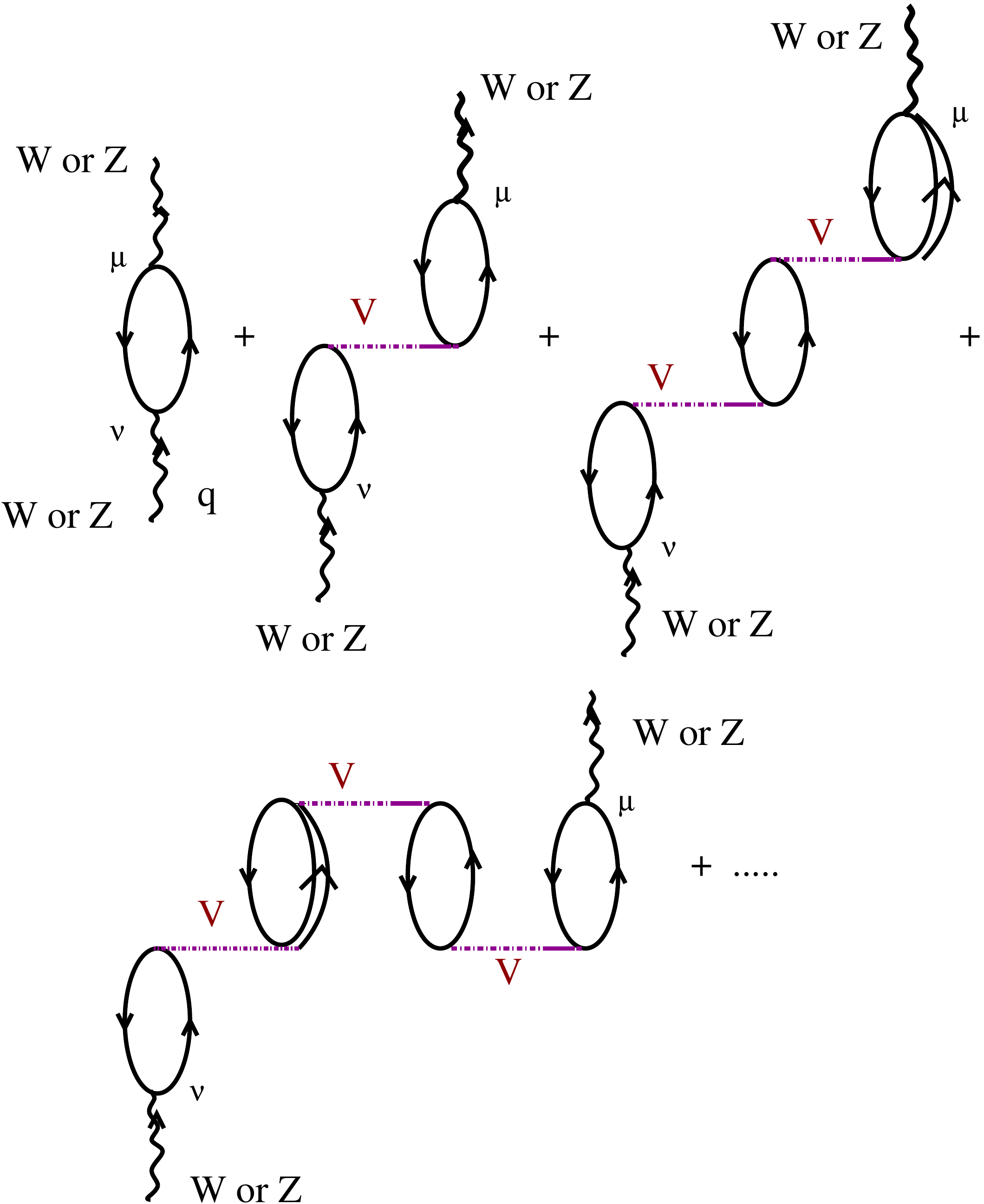}\includegraphics[height=4.0cm]{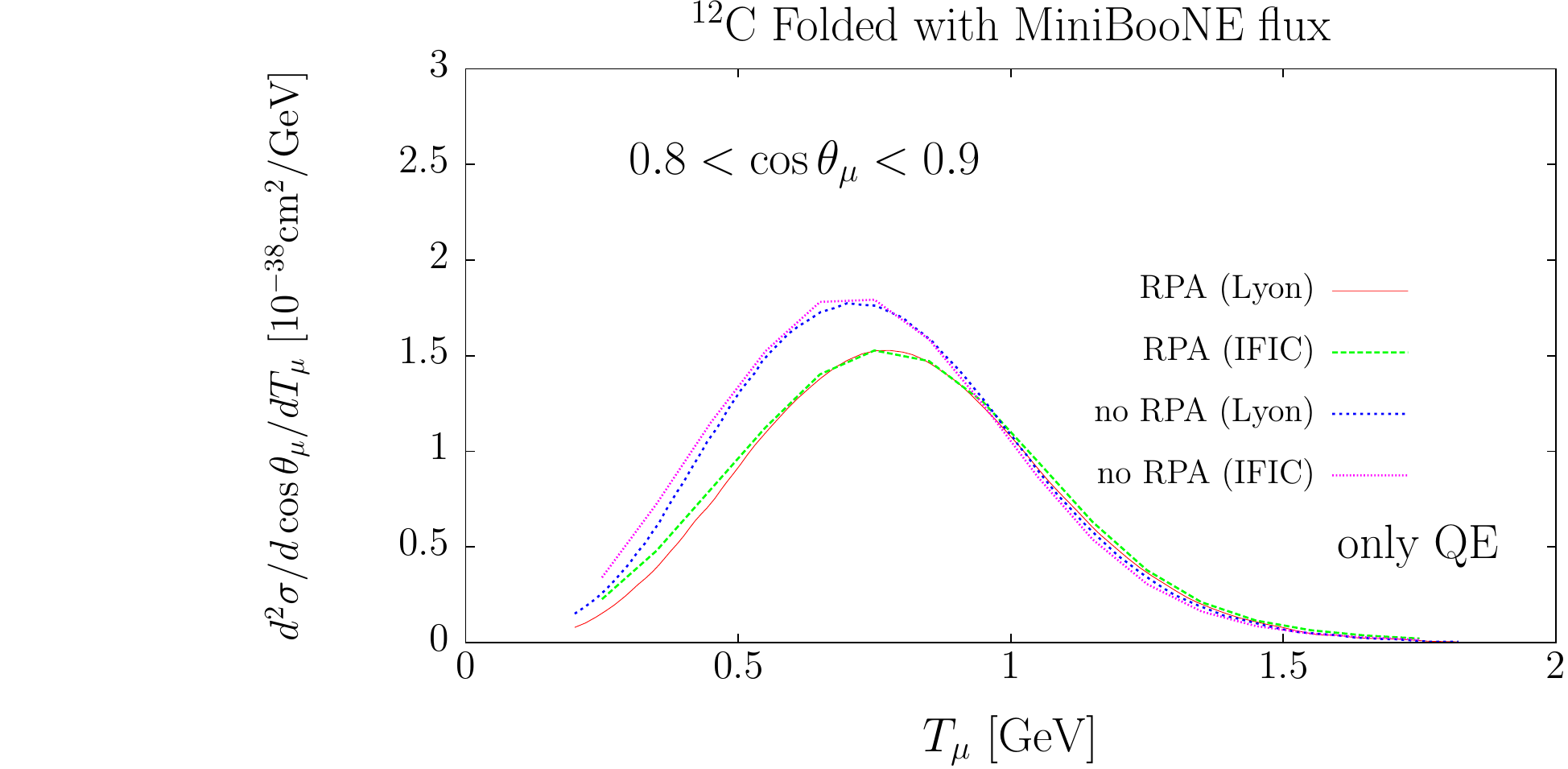}}
\\ \vspace{1cm}
\makebox[0pt]{\hspace{-1cm}\includegraphics[height=4.0cm]{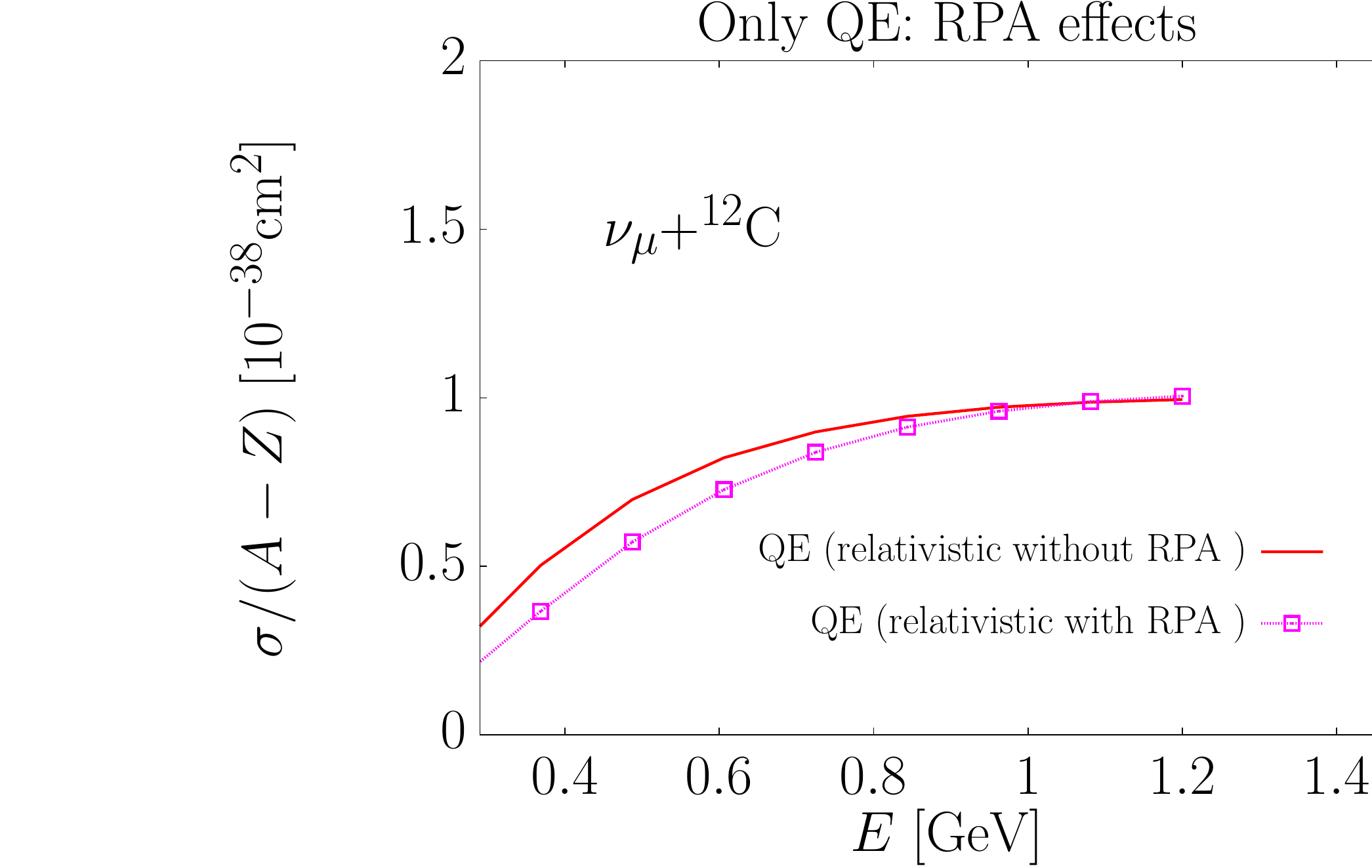}\hspace{1cm}\includegraphics[height=4.0cm]{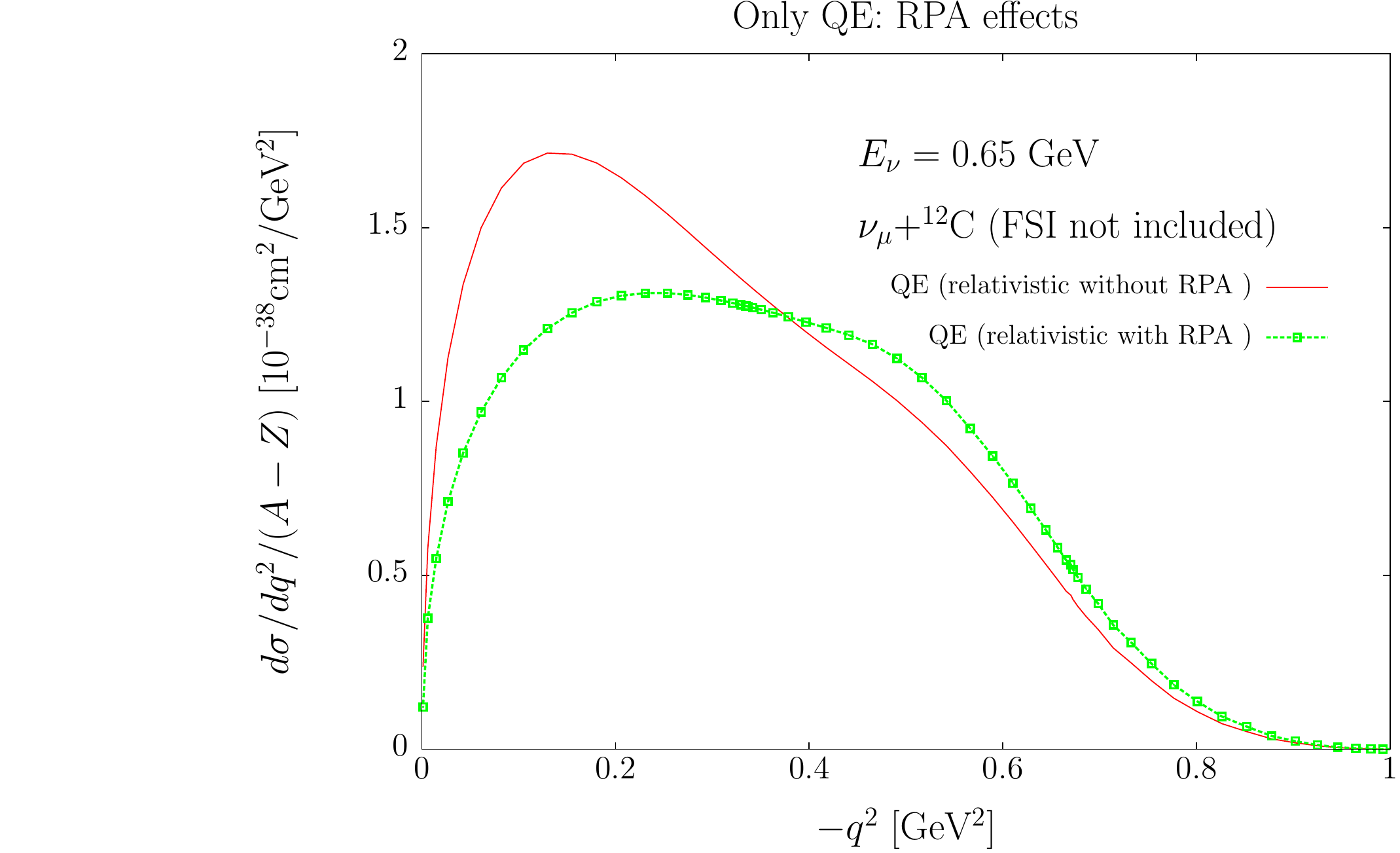}}
\end{center}
\caption{\footnotesize Top Left: Set of irreducible diagrams responsible
  for the polarization (RPA) effects in the 1p1h contribution to the
  $W$ or $Z$ self-energies. Top Right: MiniBooNE flux-averaged CC
  quasielastic $\nu_\mu- ^{12}$C double differential cross section per
  neutron for $0.8 < \cos\theta_\mu < 0.9$ as a function of the muon
  kinetic energy. Bottom: Different theoretical predictions for muon
  neutrino CCQE total cross section off $^{12}$C, as a function of the
  neutrino energy (left) and $q^2$ (right), obtained from the relativistic model of
  Ref.~\cite{23-Nieves:2004wx}.  In all cases $M_A \sim 1.05$ GeV. }\label{fig-23:fig3}
\end{figure}

Inclusion of nucleon-nucleon long-range correlations leads to RPA (Random Phase
Approximation) which improves predictions at lower momentum
transfers (and also low $Q^2$).  RPA corrections have been discussed by many authors in the past 
and recently included in computations of three groups (IFIC,
  Lyon and Aligarh\footnote{The Aligarh group uses a similar approach to that derived in \cite{23-Nieves:2004wx}, 
but with some 
simplifications that though well suited to study the related process of muon capture in nuclei, might not 
be totally appropriate for the case of larger energies and momenta being transferred 
to the nucleus (see the discussion in
\cite{23-Nieves:2004wx}).}) in Refs.~\cite{23-Nieves:2004wx,23-Nieves:2005rq},
  \cite{23-Martini:2009uj,23-Martini:2010ex}, and \cite{23-aligarh} respectively. When the
  electroweak interactions take place in nuclei, the strengths of
  electroweak couplings may change from their free nucleon values due
  to the presence of strongly interacting nucleons. Indeed, since the
  nuclear experiments on $\beta$ decay in the early 1970s \cite{23-beta},
  the quenching of axial current is a well-established phenomenon. The
  RPA re-summation accounts for the medium polarization effects in the
  1p1h contribution (Fig.~\ref{fig-23:fig2}(a)) to the $W$ and $Z$
  selfenergy by substituting it by a collective response as shown
  diagrammatically in the top left panel of
  Fig.~\ref{fig-23:fig3}. Evaluating these effects, requires an in-medium baryon-baryon
  effective force, which in both sets (IFIC and Lyon) of
  calculations was successfully used/tested in previous works on
  inclusive nuclear electron scattering.   RPA effects are important as
  can be appreciated in the top right panel of Fig.~\ref{fig-23:fig3}. In
  this plot, we show results from both IFIC and Lyon models, presented
  in Refs.~\cite{23-Nieves:2011yp} and ~\cite{23-Martini:2011wp}, respectively
  for the CC quasielastic $\nu_\mu- ^{12}$C double differential
  cross sections convoluted with the MiniBooNE
  flux~\cite{23-AguilarArevalo:2008yp}. There, we also see that
  predictions of both groups for these genuine QE contribution,
  with and without RPA effects, turn out to be in a quite good
  agreement. Finally, it is important to stress also that  RPA corrections
  strongly decrease as the neutrino energy increases, while
its effects should account for a low $Q^2$ deficit of CCQE events reported by
several experimental groups (see bottom panels of Fig.~\ref{fig-23:fig3}). 
Continuum RPA (CRPA) computations for neutrino scattering were performed by the Ghent group \cite{23-crpa}.
\subsubsection{Other theoretical developments}
In~\cite{23-Jachowicz:2007ek,23-Maieron:2003df,23-Amaro:2011qb}
the bound-state wave functions are described as self-consistent
   Dirac-Hartree solutions, derived within a relativistic mean field
  approach by using a Lagrangian containing $\sigma$ and $\omega$
  mesons~\cite{23-sw}. This scheme also accounts for some SF effects. Moreover,
  these models also incorporate the FSI
  between the ejected nucleon and the residual nucleus. The final
  nucleon is described either, as a scattering solution of the Dirac
  equation~\cite{23-Maieron:2003df,23-Amaro:2011qb} in the presence of the
  same relativistic nuclear mean field potential applied to the
  initial nucleon, or adopting a relativistic multiple-scattering
  Glauber approach~\cite{23-Jachowicz:2007ek}. 

  The relativistic Green$^\prime$s
  function model~\cite{23-Meucci:2011ce} would be also appropriate to
  account for FSI effects  between the ejected nucleon and the
  residual nucleus for the inclusive scattering, where only the
  outgoing lepton is detected. There, final-state channels are
  included, and the flux lost in each channel is recovered in the
  other channels just by the imaginary part of an empirical optical
  potential and the total flux is thus conserved. 

Another interesting approach starts with a phenomenological model
  for the neutrino interactions with nuclei that is based on the
  superscaling behavior of electron scattering data. Analysis of
  inclusive $(e, e^\prime)$ data have demonstrated that for momentum transfers 
$q>\sim 500$~MeV/c at energy
  transfers below the QE peak superscaling is fulfilled rather
  well~\cite{23-super}. The general procedure consist on dividing the
  experimental $(e, e^\prime)$ cross section by an appropriate
  single-nucleon cross section to obtain the experimental scaling
  function, which is then plotted as a function of a certain scaling
  variable for several kinematics and for several nuclei. If the
  results do not depend on the momentum transfer $q$, then
  scaling of the first kind occurs, if there is no dependence on the
  nuclear species, one has scaling of the second kind. The
  simultaneous occurrence of scaling of both kinds is called
  superscaling. The superscaling property is exact in the RFG models,
  and it has been tested in more realistic models of the $(e,
  e^\prime)$ reaction. The Super-Scaling approach (SuSA) is based on
  the assumed universality of the scaling function for electromagnetic
  and weak interactions~\cite{23-Amaro:2004bs}.  The scaling function thus
  determined from $(e, e^\prime)$ data is then directly taken over to
  neutrino interactions~\cite{23-Amaro:2004bs,23-Amaro:2006pr}. There are no
  RPA correlations or SF corrections explicitly taken into account,
  but they may be contained in the scaling function. Nevertheless,
  such approach is far from being microscopic. Moreover, it is difficult to
  estimate its theoretical uncertainties, as for example to what
  extent the quenching of the axial current, that is due to RPA
  corrections, is accounted for by means of scaling functions determined in 
  $(e, e^\prime)$ experiments, which are driven by the vector current.
\subsubsection{Theretical models versus MiniBooNE 2D data}
The MiniBooNE
data~\cite{23-AguilarArevalo:2010zc} have been quite surprising.  Firstly,
the absolute values of the cross section are too large as compared to
the consensus of theoretical models \cite{23-Boyd:2009zz,23-AlvarezRuso:2010ia}. Actually, the
cross section per nucleon on $^{12}$C is clearly larger than for free
nucleons.  Secondly, their fit to the shape (excluding normalization) of
the $Q^2$ distribution done within the RFG model leads to the axial mass, $M_A=1.35\pm 0.17$ GeV,
much larger than the previous world
average ($\approx 1.03$ GeV)
\cite{23-Bernard:2001rs,23-Lyubushkin:2008pe}. Similar results 
have been later obtained analyzing MiniBooNE data with
more sophisticated treatments of the nuclear effects that work well in
the study of electron scattering. For instance,
Refs.~\cite{23-Benhar:2009wi,23-Benhar:2010nx} using the impulse
approximation with state of the art spectral functions for the
nucleons fail to reproduce data with standard values of $M_A$.  Large
axial mass values have also been obtained in
ref.~\cite{23-Juszczak:2010ve} where the 2D differential cross section was analyzed for the first time
using RFG model and spectral
function. Similar results were obtained in Ref.~\cite{23-Butkevich:2010cr}, where the data have been
analyzed in a relativistic distorted-wave impulse approximation supplemented
with a RFG model. 
\subsection{Multinucleon mechanisms}
\label{sec:23-mnm}
A plausible solution to the large axial mass puzzle was firstly pointed out by
M. Martini\footnote{The papers of Martini et al are based on the older investigation by 
Marteau et al \cite{23-marteau}. The relavant features of the model were known already at
the end of 1990s and at that time the goal was to understand better
SupeKamiokande atmospheric neutrino oscillation signal.
} et al.~\cite{23-Martini:2009uj,23-Martini:2010ex}, and later
corroborated by the IFIC group~\cite{23-Nieves:2011yp,23-Nieves:2011pp}. 
In the MiniBooNE measurement of Ref.~\cite{23-AguilarArevalo:2010zc}, QE is
related to processes in which only a muon is detected in the final
state. As was already discussed above, besides genuine QE events, this
definition includes multinucleon processes (Fig.~\ref{fig-23:fig2}(e)\footnote{Note that
  the intermediate pion in this term is virtual and it is part of
  the $\Delta N \to NN$ interaction inside of the nucleus. Indeed, one
  should consider a full interaction model for the in medium
  baryon--baryon interaction.}), where the gauge
  boson is being absorbed by two or more nucleons, and others like real pion production followed by absorption
(Fig.~\ref{fig-23:fig2}(c) and (d)). The MiniBooNE analysis of the data
attempts to correct (through a Monte Carlo estimate) for some of these latter effects,
 such as real pion production that
escapes detection through reabsorbtion in the nucleus leading to
multinucleon emission. But, it seems clear that to describe the data
of Ref.~\cite{23-AguilarArevalo:2010zc}, it is necessary to consider, at
least, the sum of the selfenergy diagrams depicted in
Figs.~\ref{fig-23:fig2}(a) and (e). Those correspond to the genuine QE
(absorption by just one nucleon), and the multinucleon contributions,
respectively. The sum of these two
contributions contribute to the CCQE-like cross
section\footnote{Also for simplicity, we will often refer to the
  multinucleon mechanism contributions, though they include effects
  beyond gauge boson absorption by a nucleon pair, as 2p2h (two
  particle-hole) effects.}.

The inclusion of the  2p2h contributions enables ~\cite{23-Nieves:2011yp,23-Martini:2011wp} 
the double differential cross section $d^2\sigma/dE_\mu
d\cos\theta_\mu$ and the integrated flux unfolded cross
section\footnote{We should warn the reader here, because of the
  multinucleon mechanism effects, the algorithm used to reconstruct
  the neutrino energy is not adequate when dealing with
  quasielastic-like events, a distortion of the total flux
  unfolded cross section shape could be produced. We will address this
  point in Subsect.~\ref{sec-23:ereco}.}  measured by MiniBooNE, to be described with
values of $M_A$ (nucleon axial mass) around $1.03\pm 0.02$ GeV~\cite{23-Bernard:2001rs,23-Lyubushkin:2008pe}. This is re-assuring
from the theoretical point of view and more satisfactory than the
situation envisaged by some other works that described the MiniBooNE
data in terms of a larger value of $M_A$ of around 1.3--1.4 GeV, as
mentioned above. 
\subsubsection{Similarites and differences between multinucleon ejection models}
\begin{figure}[tbh]
\begin{center}
\makebox[0pt]{\hspace{-2cm}\includegraphics[height=5.0cm]{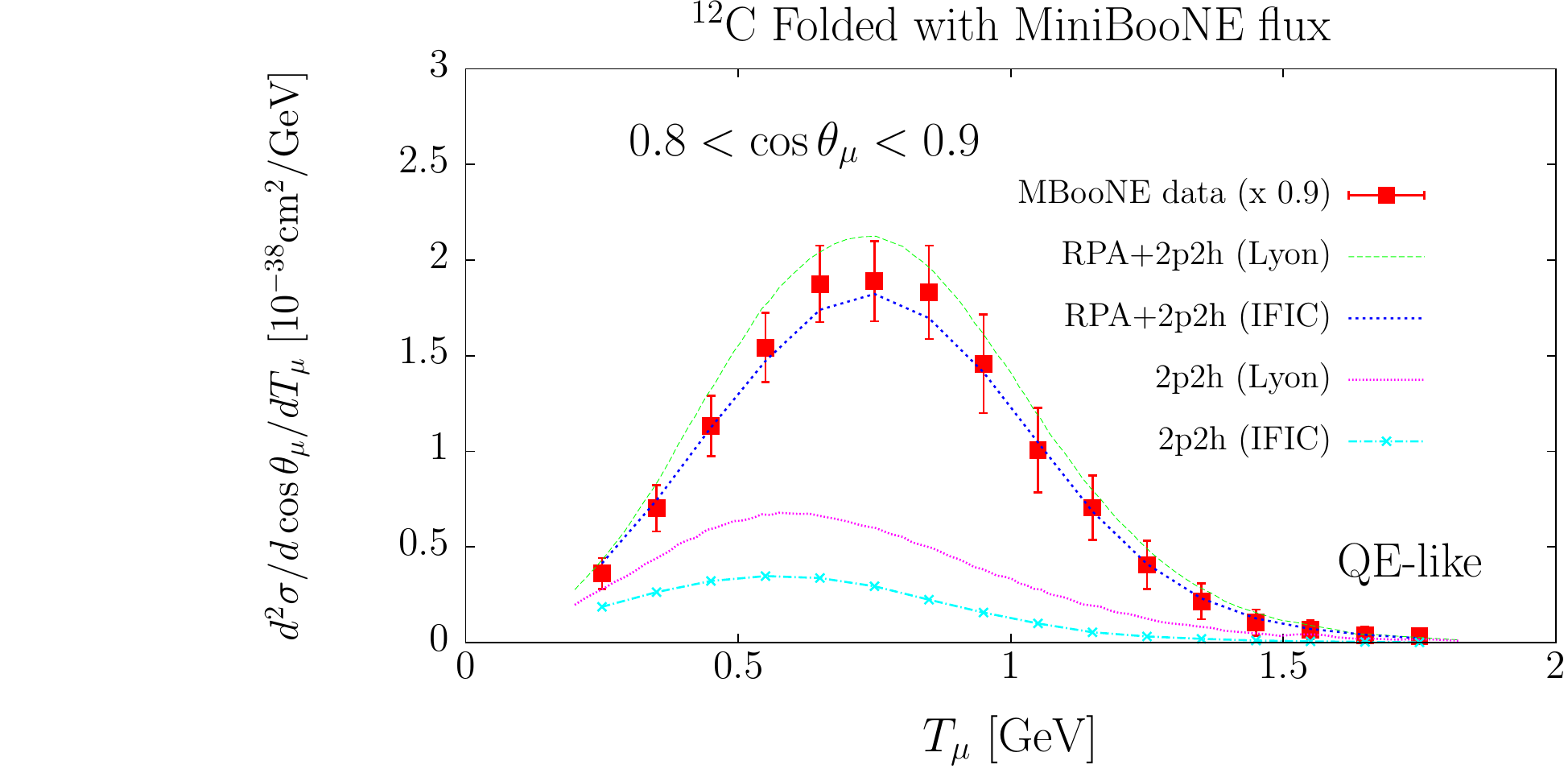}}
\caption{\footnotesize MiniBooNE flux-averaged CC
  quasielastic $\nu_\mu- ^{12}$C double differential cross section per
  neutron for $0.8 < \cos\theta_\mu < 0.9$, as a function of the muon
  kinetic energy. Experimental data from Ref.~\cite{23-AguilarArevalo:2010zc}
are multiplied by 0.9. In all the cases $M_A \sim 1.05$ GeV. }\label{fig-23:fig4}
\end{center}
\end{figure}
As shown in the top panel of Fig.~\ref{fig-23:fig3},
the IFIC group predictions~\cite{23-Nieves:2011yp,23-Nieves:2011pp} for QE
cross-sections  agree quite well
with those obtained in Refs.~\cite{23-Martini:2009uj,23-Martini:2010ex,
  23-Martini:2011wp} (Lyon group). However, both above presented approaches considerably differ (about a factor of two)
in their estimation of
the size  of the multinucleon effects, as
can be appreciated in Fig.~\ref{fig-23:fig4}. IFIC predictions, when the 2p2h contribution is
included, favor a global normalization scale of about 0.9 (see
\cite{23-Nieves:2011yp}). This is consistent with the MiniBooNE
estimate of a total normalization error as 10.7\%. The IFIC
evaluation in ~\cite{23-Nieves:2011pp,23-Nieves:2011yp}, of 
multinucleon emission contributions to the cross section is fully
microscopical and it contains terms, which were either not considered
or only approximately taken into account in \cite{23-Martini:2009uj,
  23-Martini:2010ex, 23-Martini:2011wp}. Indeed, the results of these latter
works rely on some computation of the 2p2h mechanisms for the $(e,e')$
inclusive reaction (\cite{23-Alberico:1983zg}), which results are simply used
for neutrino induced processes without modification. Thus, it is clear that these latter
calculations do not contain any information on axial or axial-vector
contributions\footnote{The evaluation of the nuclear response induced
  by these 2p2h mechanisms carried out in Ref.~\cite{23-Martini:2009uj}
  is approximated, as acknowledge there.  Only, the contributions in
  \cite{23-Martini:2009uj} that can be cast as a $\Delta-$selfenergy
  diagram should be quite similar to those derived in
  ~\cite{23-Nieves:2011pp} by the IFIC group, since in both cases the
  results of Ref.~\cite{23-Oset:1987re} for the $\Delta-$selfenergy are
used.}. For antineutrinos the IFIC model predicts, contrary
to the results of the Lyon group, also a sizeable effect of 2p2h
excitations.

Another microscopic approach to 2p2h excitations was proposed by
Amaro et al. These authors have used the empirical $(e,e^\prime)$ SuSA scaling
function to describe the CCQE MinibooNE data, including 
some 2p2h contributions due to MEC (meson exchange
currents)~\cite{23-Amaro:2010sd,23-Amaro:2011aa}. The approach, used in these
latter works, to  evaluate the 2p2h effects, though fully relativistic,
does not contain the axial contributions. The authors of~\cite{23-Amaro:2010sd,23-Amaro:2011aa} 
also find an increase of the inclusive
cross section for neutrinos; at forward muon angles the
calculations come close to the data, but the MEC  contributions
die out fast with increasing angle so that the
cross section is significantly underestimated at backward
angles. As a consequence the energy-separated (flux unfolded) cross section
obtained for the MiniBooNE experiment while being
higher than that obtained from SuSA alone still
underestimates the experimental result even when 2p2h
contributions are added. Recently, a strong difference between
neutrino and antineutrino cross sections has been
obtained within this model, with the 2p2h effects being
significantly larger for antineutrinos than for 
neutrinos~\cite{23-Amaro:2011aa}. 

Two other effective models to account for MEC/2p2h effects have been
proposed by Bodek et al.~\cite{23-bode_mec} [transverse enhancement
model (TEM)] and Lalakulich et al.~\cite{23-Lalakulich:2012ac}. The TEM
can easily be implemented in MC event generators
\cite{23-Sobczyk:2012ms}. It assumes that it is sufficient to describe
properly an enhancement of the transverse electron QE response
function keeping all other ingredients as in the free nucleon target
case. Thus, some effective proton and neutron magnetic form factors
are fitted to electron-nucleus data and later they are used, together
with the free nucleon axial current, to study CCQE processes.  It is to
say, the TEM assumes that there are no nuclear medium effects (RPA,
2p2h mechanisms, etc...)  affecting those nuclear response functions
induced by the nucleon axial-vector current. Despite of a certain
phenomenological success to describe the MiniBooNE data~\cite{23-bode_mec,
23-Sobczyk:2012ms}, such assumption seems quite unjustified.

In the model of Ref.~\cite{23-Lalakulich:2012ac}, the multinucleon
mechanism contributions are parametrized as phase space multiplied by
a constant which is fitted to the difference of the energy-separated
MiniBooNE data and the calculated QE cross section. RPA effects are
not taken into account in ~\cite{23-Lalakulich:2012ac}. Since these
tend to lower the cross section in particular at forward muon angles,
the model of \cite{23-Lalakulich:2012ac} underestimates the
contributions of 2p2h effects there. Indeed, the authors of this
reference find that the shape and over-all size of the 2p2h
contribution turns out to be rather independent of the muon
angle. This is in sharp contrast with the microscopical
results obtained within the
IFIC~\cite{23-Nieves:2011pp,23-Nieves:2011yp} and SuSa
models~\cite{23-Amaro:2011aa}, that find the 2p2h contribution becomes
significantly less important as the muon scattering angle increases.
\subsubsection{Perspectives to measure the MEC/2p2h contribution}
The unambiguous experimental measurement of the MEC contribution to the CC inclusive cross section
can be made by detecting hadrons in the final state. All the microscopic models provide up to now only the 
MEC/2p2h contribution to the muon inclusive
2D differential cross section: $d^2\sigma_{\nu_\ell
    \ell}/d\Omega(\hat{k^\prime})dE^\prime_\ell$. Such models cannot
  describe detailed exclusive cross sections (looking into the nucleon
  side), as explicit FSI effects, that modify the outgoing nucleon spectra, 
have not been addressed yet in these microscopical models. It is reasonable to assume
that at the level of the primary reaction mechanism, they produce only
  slightly changes in $d^2\sigma_{\nu_\ell
    \ell}/d\Omega(\hat{k^\prime})dE^\prime_\ell$, leaving almost
  unchanged the integrated cross
  sections~\cite{23-Benhar:2006nr,23-Ankowski:2007uy}.

A model to describe hadrons in
the final state was proposed in \cite{23-Sobczyk:2012ms}. It was implemented in the NuWro MC event generators
and its predictions were used in the analysis of recent MINERvA antineutrino CCQE data.

In the papers \cite{23-Lalakulich:2012ac,23-Sobczyk:2012ms} various observables are discussed which can be used to detect MEC contribution.
One option is to look at proton pairs in the final state. Another possibility is to investigate
the distribution of visible energy which allows to include contributions from protons below reconstruction threshold.
The basic intuition from the electron scattering is that MEC events populate the region between QE and $\Delta$ peaks. 
Typically,
to have a MEC event more energy must be transfered to the hadronic system than for a CCQE one. 
However, it should be stressed that the precision with which FSI effects are currently handled in MC
codes can make such a measurement difficult. During last few years FSI studies 
were focused on pions only \cite{23-pion_fsi_neut} aiming at understanding recent pion production
data on nuclear targets \cite{23-nuwro_fsi}. Nucleons in the final state were never studied with
a similar precision so there is less data to benchmark nucleon FSI effects.
\subsection{Monte Carlo event generators}
Monte Carlo codes (GENIE, NuWro, Neut,
Nuance, etc) describe CCQE events using a simple RFG
model, with FSI effects implemented by means of
a semi-classical intranuclear cascade. NuWro offers also a possibility to run simulations with
spectral function and an effective momentum dependent nuclear potential. It is also by now the only
MC generator with implementation of MEC dynamics. Since the primary  interaction and 
the final state effects are effectively decoupled, FSI do not change
the total and outgoing lepton differential cross sections. 
\subsection{Neutrino energy reconstruction}
\label{sec-23:ereco}
Neutrino oscillation probabilities depend
on the neutrino energy, unknown for broad fluxes and
often estimated from the measured angle and energy
of the outgoing charged lepton $\ell$ only. This is the situation of the experiments 
with  Cherenkov detectors where protons in the final state 
are usually below the Cherenkov threshold. Then, it is common to define a 
reconstructed neutrino energy $\ereco$ (neglecting binding energy and the difference of proton and 
neutron masses) as:
\begin{equation}
\ereco = \frac{M
  E_\ell-m_\ell^2/2}{M-E_\ell+|\vec{p}_\ell|\cos\theta_\ell}\label{eq-23:defereco}
\end{equation}
which would correspond to the energy of a neutrino that emits a lepton,
of energy $E_\ell$ and three-momentum $\vec{p}_\ell$, with a gauge boson
$W$ being absorbed by a free nucleon of mass $M$ {\em at rest} in a CCQE event. 
Each event contributing to the flux
averaged double differential cross section $d\sigma/dE_\ell
d\cos\theta_\ell$ defines unambiguously a value of $\ereco$.  The
actual (``true'') energy, $E$, of the neutrino that has produced the
event will not be exactly $\ereco$.  Actually, for each $\ereco$,
there exists a distribution of true neutrino energies that give
rise to events whose muon kinematics would lead to the given value of
$\ereco$. In the case of genuine QE events, this distribution is sufficiently 
peaked (the Fermi motion broadens the peak and binding energy shifts it a little) around the
true neutrino energy to make the algorithm in 
Eq.~(\ref{eq-23:defereco}) accurate enough to study the neutrino
oscillation phenomenon~\cite{23-Meloni:2012fq} or to extract neutrino
flux unfolded CCQE cross sections from data (assuming that  the neutrino flux
spectrum is known)~\cite{23-Martini:2012fa,23-Nieves:2012yz}.  The effect of this assumption on
the much more demanding measurement of CP-violation effects is currently being evaluated.

However, and due to presence of multinucleon events, there is a long tail in the
distribution of true energies associated to each $\ereco$ that makes
the use of Eq.~(\ref{eq-23:defereco}) unreliable. The effects of the inclusion of multinucleon
processes on the energy reconstruction have been noticed in \cite{23-Sobczyk:2012ms} and investigated in
Ref.~\cite{23-Martini:2012fa}, within the Lyon 2p2h model and also estimated
in Ref.~\cite{23-Mosel:2012hr}, using the simplified model of
Ref.~\cite{23-Lalakulich:2012ac} for the multinucleon mechanisms. This
issue has been more recently also addressed in the context of the IFIC
2p2h model in Ref.~\cite{23-Nieves:2012yz}, finding results in a qualitative
agreement with those of Refs.~\cite{23-Martini:2012fa} and
\cite{23-Mosel:2012hr}. 

In Ref.~\cite{23-Nieves:2012yz} it is also studied in detail the
$^{12}$C unfolded cross section published in
\cite{23-AguilarArevalo:2010zc}. It is shown there that the unfolding procedure is
model dependent. Moreover, it is also shown that the MiniBooNE published CCQE cross section as a function 
of neutrino energy
differs from the real one. This is because the MiniBooNE analysis assumes that all the events are
QE. The authors of
\cite{23-Nieves:2012yz} finally conclude that the MiniBooNE unfolded cross section
exhibits an excess (deficit) of low (high) energy neutrinos, which is mostly 
an artifact of the unfolding process that ignores multinucleon
mechanisms.
\subsection{NC elastic}
MiniBooNE has also measured flux integrated NC elastic reaction cross-section \cite{23-mb_elastic}. Using these data,
the best fit value of the axial mass was found te be $M_A=1.39\pm0.11$~GeV. The measurement was possible 
because the 
MiniBooNE Cherenkov detector can observe also scintillation light from low 
momentum nucleons. 
An attempt was done to measure the nucleon strange quark component using the proton enriched sample of events with
a result consistent with zero: $\Delta s=0.08\pm 0.26$.
\subsubsection{Theoretical considerations}
The MiniBooNE NCEl data were analyzed in~\cite{23-ncel_butkevich}. The fit was done to the $Q^2$ distribution of events with 
the best fit value of $M_A$ equal to $1.28\pm 0.05$~GeV. Moreover the authors of~\cite{23-ncel_benhar} concluded
that axial mass as large as $1.6$~GeV is still too small to reproduce the MiniBooNE NCEl data. Critical discusson 
of this statement can be found in Ref.~\cite{23-ncel_ankowski}.
\section{The Resonance  Region}
In the RES region the degrees of freedom are hadronic resonances, 
with the most important being the $\Delta (1232)$. 
Typical final states are those with a single pion. During the last five years several new pion production measurements
have been performed. In all of them the targets were nuclei (most often carbon) and interpretation of the data 
in terms of the
neutrino-nucleon cross section needed to account for nuclear effects, impossible to do in a model independent manner. 
Because of that
it has become a standard that the published data include nuclear effects with most uncertain FSI.
Perhaps not surprisingly, in several papers old deuterium ANL and BNL pion production
data were re-analyzed aiming to better understand the pion production reaction on free nucleons. Theoretical
models became more sophisticated and the major improvement was a development of well justified mechanisms
for the non-resonant contribution in the $\Delta$ region. 
Some papers addressed the problem of higher resonances, a topic which will be investigated
experimentally with future MINERvA results.
On the other hand, there has been a lot of activity in the area of the coherent pion production and this subject will be discussed
separately.
\subsection{Experimental Results}
\subsubsection{NC$\pi^0$}
Neutral current $\pi^0$ production (NC$\pi^0$) is a background to $\nu_e$ appearance oscillation signal.  One is interested in a $\pi^0$ leaving the nucleus and 
recent experimental data are given in this format with all the FSI effects included. 
Signal events originate mostly from: NC$1\pi^0$ primary interaction with a $\pi^0$ not being affected by FSI and
NC$1\pi^+$ primary interaction with the $\pi^+$ being transformed into $\pi^0$ in a charge exchange FSI reaction.
An additional difficulty in interpreting the NC$\pi^0$ production comes from a coherent (COH) contribution.
In the case of MiniBooNE flux neutrino-carbon reactions ($<E_\nu> \sim 1$~GeV)  it is estimated to account for 
$\sim~20$\% of signal events \cite{23-mb_coh}.

Four recent measurements of NC$\pi^0$ production (K2K \cite{23-k2k_ncpi0}, MiniBooNE neutrinos, MiniBooNE 
antineutrinos \cite{23-mb_ncpi0}, 
SciBooNE \cite{23-sb_ncpi0}) are complementary. They use three different fluxes: (K2K, Fermilab Booster 
neutrinos and anti-neutrinos) and 
three targets: $H_2O$ (K2K), $CH_2$ (MiniBooNE) and $C_8H_8$ (SciBooNE). MiniBooNE presented the results 
in the form of absolutely 
normalized cross-section while K2K and SciBooNE reported only the ratios $\sigma (NC1\pi^0)/ \sigma (CC)$. 
There is an important
difference in what was actually measured: K2K and MiniBooNE present their results as measurements of final 
states with only one 
$\pi^0$ and no other mesons. SciBooNE defines the signal as states with {\it at least} one $\pi^0$ in the final 
state so that a contamination from 
$1\pi^01\pi^\pm$, $2\pi^0$ and $>2\pi$ (with $>1\pi^0$) final states is included and its fraction can be
estimated to be 17\% \cite{23-nuwro_fsi}. 
Final results are presented as flux averaged distributions of events 
as a function of the $\pi^0$ momentum, 
and in the case of MiniBooNE and SciBooNE also as a function of the $\pi^0$ production angle. 
\subsubsection{CC$\pi^+$}
\begin{figure}[htb]
\begin{center}
\makebox[0pt]{\includegraphics[height=6.0cm]{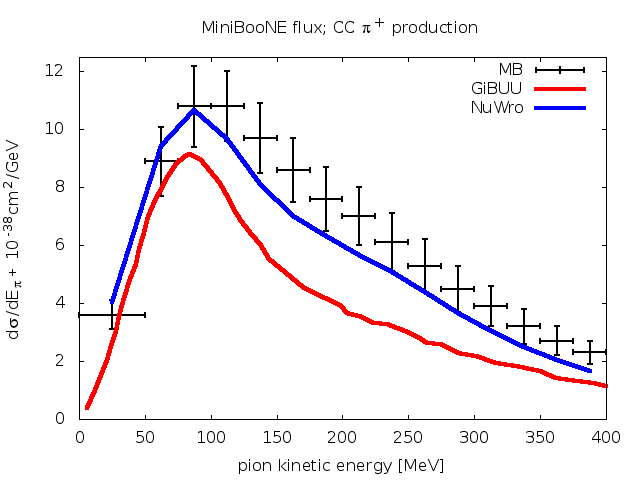}\includegraphics[height=6.0cm]{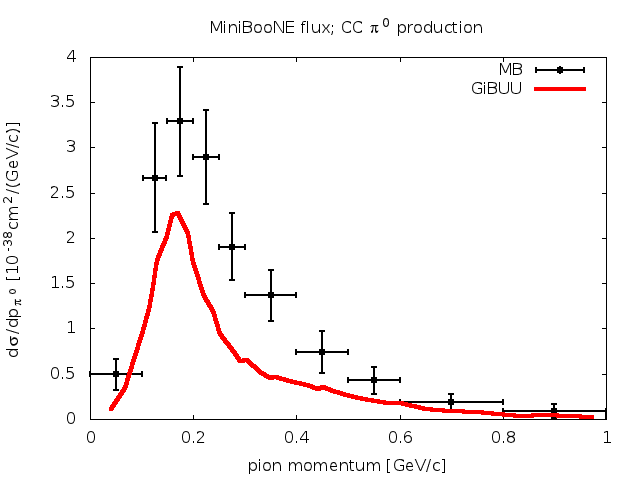}}
\end{center}
\caption{\footnotesize Left: Differential cross section for CC$1\pi^+$ production in the final state (all the FSI effects are 
included). MiniBooNE measurement  \cite{23-mb_ccpiplus} is compared to GiBUU \cite{23-gibuu_pions1} and NuWro computations.
Right: the same for CC$\pi^0$ production, but only GIBUU results are shown. }\label{fig-23:pions}
\end{figure}

MiniBooNE measured CC $1\pi^+$ production cross sections, where the signal is defined as exactly one $\pi^+$ in the final
state with no other mesons \cite{23-mb_ccpiplus}. A variety of flux integrated differential cross sections, 
often double differential were
reported in $Q^2$ and final state particles momenta. Also absolute $\pi^+$ production cross sections as a function of 
neutrino energy are provided in Ref.~\cite{23-mb_ccpiplus}. The cross section results are much larger than NUANCE MC predictions
and the difference is on average $23\%$. In Fig. \ref{fig-23:pions} on the left GiBUU and NuWro predictions for CC$\pi^+$ are compared to the MiniBooNE
data. 
\subsubsection{CC$\pi^0$}
MiniBooNE measured also CC $1\pi^0$ production cross sections. As before, the signal is defined as exactly one $\pi^0$ in the final
state \cite{23-mb_ccpi0}. Various differential distributions are available. There is a dramatic discrepancy 
between the measured CC $1\pi^0$
production cross section as a function of neutrino energy and NUANCE MC predictions in the region of lower 
energies.
On average the data is larger by $56\pm 20$\%, but for $E_\nu<1$~GeV the disagreement is as large as a factor of $2$.
In Fig. \ref{fig-23:pions} on the right GiBUU predictions for CC$\pi^+$ are compared to the MiniBooNE
data. 
\subsubsection{Ratio $\sigma (CC1\pi^+)/\sigma (CCQE)$}
Another useful MiniBooNE measurement was the ratio $\sigma (CC1\pi^+)/\sigma (CCQE)$ \cite{23-mb_ratio}. The ratio of 
CC1$\pi^+$-like (one pion in the final state) to CCQE-like cross-sections on $CH_2$ as a function of neutrino energy
was measured with an accuracy of $\sim 10$\% in bins with highest statistics. This measurement puts constraints
on the theoretical models which include QE, $\Delta$ excitation and MEC/2p2h dynamics. But still, in order 
to compare with theoretical model predictions to these data, 
FSI effects must be included. In order to make such a comparison easier, MiniBooNE provided also 
FSI corrected data
representing the ratio of CC$1\pi^+$/CCQE cross sections at the primary interaction. The corrected results 
are biased by MC assumptions and in particular they neglect most of the MEC/2p2h contributions which is
contained in the QE-like sample of events. Finally, MiniBooNE re-scaled their results in order to get data points
for an isoscalar target and enable comparison to old ANL and also more recent K2K data~\cite{23-k2k_ratio}. 
K2K measured ratio of cross sections on bound nucleons inside the nucleus corrected for FSI effects.
CC$1\pi^+$ events were not identified on an event-by-event basis.
\subsubsection{Theoretical Considerations}
Due to nuclear effects a comparison to the new data is possible only for MC event generators,
sophisticated computation tools like GiBUU and also a few theoretical groups which are able to evaluate FSI
effects.

Most of the interesting work was done within GiBUU. It turned out to be very difficult reproduce 
the MiniBooNE CC$1\pi^+$ and CC$1\pi^0$ results: the measured cross section is much larger than theoretical
computations. In the case of CC $1\pi^+$ production the discrepancy is as large as 100\%. It was also noted
that the reported shape of the distribution of $\pi^+$ kinetic energies is different from theoretical calculations
and does not show a strong decrease at $T_{\pi^+}>120$~MeV located in the region of maximal probability for 
pion absorption. 

The authors of \cite{23-gibuu_pions1} mention three possible reasons for the data/GiBUU predictions discrepancy: 
(i) the fact that 
$\Delta$ excitation axial form factor was chosen to agree with the ANL data only, neglecting the larger cross section
BNL measurements; (ii) hypothetical 2p-2h-1$\pi$ pion production contribution analogous to 2p-2h discussed
in the Sect. \ref{sec:23-mnm}; (iii)
flux  underestimation in the MiniBooNE experiment. For the last point, the argument gets support from the better 
data/theory agreement found for the ratio, as discussed below.

In the case of NC$\pi^0$ production, a systematical comparison was done with NuWro MC predictions 
with an updated FSI model
for pions \cite{23-nuwro_fsi}. The overall agreement is satisfactory. Shapes of 
the distributions of final state $\pi^0$'s
are affected by an interplay between pion FSI such as absorption and {\it formation time} effects, 
understood here as an effect
of a finite $\Delta$ life-time. It is argued that NC$\pi^0$ production data can be very useful for benchmarking 
neutrino MC event generators.

Because of the apparent data/MC normalization discrepancy for the CC $\pi^+$ production, the interesting
data is that for the ratio $\sigma (CC1\pi^+-like)/\sigma (CCQE-like)$. This observable is free from the overall flux
normalization uncertainty. However, it is not a direct observable quantity because in the experimental analysis
it is necessary to reconstruct the neutrino energy and the procedures applied 
for the denominator and numerator are different.
Three theoretical predictions for the ratio were published. The Giessen group compared to the MiniBooNE 
ratio data using the model  described in \cite{23-leitner_luis_mosel} with the 
FSI effects modeled by the GiBUU code \cite{23-gibuu_ratio}. There is a significant discrepancy between the model and the 
data points: the calculated ratio is smaller. For the K2K data, the GiBUU model computations are 
consistent with the experimental results.

The $\sigma (CC1\pi^+)/\sigma (CCQE)$ ratio was also analyzed in Ref.~\cite{athar_ratio}. 
In this analysis many nuclear effects were included: the in medium $\Delta$ self-energy (both real and imaginary parts),
FSI effects within the cascade model of Ref.~\cite{mashnik}, RPA corrections for the CCQE... 
Computations did not include contributons from the non-resonant background and from higher resonances. The 
contribution from the coherent pion production evaluated with the model of Ref.~\cite{23-singh_coh} 
(about 5\% of the $\pi^+$ signal, a surprisingly large fraction) 
was also included in computations. The model predictions 
agree with MiniBooNE measurement for $E_\nu<1$~GeV and are below MiniBooNE data for larger neutrino energies. 

Finally, NuWro MC results for the ratio given in Ref.~\cite{23-wro_war} are slightly below the data points
for larger neutrino energies.
\subsection{Theoretical Analyses}
It has been known since ANL and BNL pion production measurements 
that although being a dominant mechanism, $\Delta$ excitation alone cannot reproduce 
the data and that nonresonant background terms must be included in the theoretical models. 
There were many attempts in the past to develop suitable models
but usually they were not very well justified from the theoretical point of view. 
\subsubsection{Nonresonant background}
A general scheme to analyze weak pion production in the $\Delta$ region based on the chiral symmetry 
was proposed a few years ago in \cite{23-nieves_chft}. 
The model is supposed to work well in the kinematical
region $W<1.3-1.4$~GeV i.e. in the $\Delta$ region. The background contribution is particulary
important at the pion production threshold, for values of $W$ near $M+m_\pi$. 
Vector form factors are taken from 
the electroproduction data and fits to helicity
amplitudes~\cite{23-olga1}. 
Although particulary important 
for the channels $\nu_\ell n\rightarrow \ell^- p\pi^0$ and $\nu_\ell n\rightarrow \ell^- n\pi^+$ the background terms contribute
also to the channel $\nu_\ell p\rightarrow \ell^- p\pi^+$ changing the fitted values of the nucleon-$\Delta$ transition 
matrix elements. A comparison to existing NC pion production data was done as well and
a good agreement was also found. An interesting question raised by the authors of \cite{23-nieves_chft} 
is that of unitarity. Their
approach does not satisfy requirements of the 
Watson theorem and this can have some consequences e.g. worse agreement with the 
antineutrino pion production data. 

The model of the nonresonant background was used by the Giessen group which made several qualitative 
comparisons to both the ANL
and BNL pion production data in the region $W<1.4$~GeV neglecting deuterium effects \cite{23-olga_anl_bnl}. 
In the case of neutron
channels the model predictions are much below the BNL data points and this is because the axial form factor
parameters were optimized to the ANL data only. This choice goes back to the 
paper \cite{23-olga1} where the authors came to the conclusion that 
the ANL and BNL data for the $\Delta^{++}$ excitation are not compatible. 
\subsubsection{Reanalysis of old bubble chamber data}
The issue of nucleon-$\Delta$ transition matrix element was 
discussed
also in other papers. The questions are: what is 
the value of the $C^A_5(0)$? How relevant are deuterium
nuclear effects in dealing with ANL and BNL data? How much tension is there between both data samples?

In Ref.~\cite{23-nieves_chft} a fit was done to the ANL data in the $\Delta^{++}$ channel 
only with the results: $C^A_5(0)=0.867\pm 0.075$ 
and $M_{A\Delta}=0.985\pm 0.082$~GeV. The obtained value of $C^A_5(0)$ was very
different from what follows from off-diagonal Goldberger-Treiman relations
($C^A_5(0)\approx 1.15$). 

The authors of \cite{23-wro_war} made a fit to both ANL and BNL
data including in the $\chi^2$, terms with 
the overall flux normalization uncertainties, separate for ANL and BNL. 
In the fit the deuterium nuclear effects were included as correction factors to the $Q^2$
distributions of events, using the results of \cite{23-singh_deu}. 
The main conclusion was that ANL and BNL data are in fact consistent. 
This statement was verified in a rigorous way using parameter goodness of fit method \cite{23-goodness}.
In the dipole parameterization 
of the $C^A_5(Q^2)$ 
form factor the best fit values were found to be $C^A_5(0)=1.14\pm 0.08$ and $M_A=0.95\pm 0.04$. 
Only $\Delta^{++}$ channel was analyzed and like in 
Ref.~\cite{23-leitner_luis_mosel} non-resonant background contributions was not included.

So far the most complete analysis of both ANL and BNL data 
was performed in \cite{23-nieves_deut}: a nonresonant background was
included and also deuterium effects were taken into account in the systematic way.
The authors made several fits with various assumptions (see Table I) {\bf and} in  the fit IV they obtained 
$C^5_A(0)=1.00\pm 0.11$. 
\subsubsection{Other theoretical approaches}
In Ref.~\cite{23-satolee1} the dynamical pion cloud effects are imposed on bare quark 
$N-\Delta$ transition matrix elements.
The model is able to reproduce both ANL and BNL weak pion production data.

The authors of \cite{23-barbero} focus on the consistent use of the $\Delta$ propagator. 
They show that the computations
relying on the standard Rarita-Schwinger propagator could lead to an underestimation of the 
weak pion production cross section. 
\subsection{Coherent pion production}
In coherent pion production (COH) the target nucleus remains in the ground state. There are four possible channels, 
for CC and NC reactions, neutrinos and anti-neutrinos. 
A clear experimental signal for the COH reaction for high energies was observed and the aim of recent measurements 
was to fill a gap in the 
knowledge of a region around $\sim 1~GeV$ COH cross-sections. At larger neutrino energies a recent 
measurement was done by 
MINOS which reported a 
NC reaction cross section at $<E_\nu>=4.9$~GeV to be consistent with the predictions of the Berger-Sehgal model 
(see below).
\subsubsection{Experimental Results}
In the case of the NC reaction, MiniBooNE \cite{23-mb_coh} and SciBooNE \cite{23-sb_coh_nc} 
searched for the COH component. SciBooNE \cite{23-sb_coh_nc} evaluated the ratio 
of the COH NC$\pi^0$ 
production to the total CC cross-section as $(1.16\pm 0.24)\%$. 

For the NC reaction 
MiniBooNE evaluated the COH component (plus possible hydrogen diffractive contribution about which 
little is known) in 
the NC$\pi^0$ production as 19.5\% (at $<E_\nu>\sim 1$~GeV) and then the overall flux averaged overall NC$1\pi^0$ 
cross-section as 
$(4.76\pm0.05\pm0.76)\cdot 10^{-40}$cm$^2$/nucleon. Unfortunately, it is difficult to translate both measurements 
into the 
absolutely normalized value of the NC COH cross-section because of strong
dependence on the NUANCE MC generator used in the data analysis.  
In NUANCE, RES, COH and BGR (nonresonance background) NC$\pi^0$ reactions are defined according to primary 
interaction and
COH pions are also subject to FSI. 
In the MiniBooNE analysis the fit is done for the composition of the sample of NC$\pi^0$ events in terms of 
three components, 
and the COH fraction is defined as $x_{COH}/(x_{COH}+x_{RES})$. 

In the case of the CC reaction, K2K \cite{23-k2k_coh} and SciBooNE 
\cite{23-sb_coh_cc} reported 
no evidence for the COH component. 
For the K2K analysis, 
the 90\% confidence limit upper bound for the COH cross-sections on carbon was estimated to be $0.6\%$ of the inclusive 
CC cross-section.
The SciBooNE upper limits (also for the carbon target) are: $0.67\%$ at $<E_\nu>\sim 1.1$~GeV, and $1.36\%$ at 
$<E_\nu>\sim 2.2$~GeV.
SciBooNE reported also the measurement of the ratio of CC COH $\pi^+$ to NC COH $\pi^0$ production and estimated it 
as $0.14^{+0.30}_{-0.28}$. This is a surprisingly low value, which disagrees with results from the 
theoretical models which at SciBooNE energies typically predict values somehow smaller $2$. For massless
charged leptons isospin symmetry implies the value of $2$ for this ratio and the finite mass corrections 
make the predicted 
ratio smaller.
\subsubsection{Theoretical developments}
Higher neutrino energy ($E_\nu >\sim 2$~GeV) COH production data (including recent NOMAD measurement) were successfully 
explained with 
a PCAC based model \cite{23-rs_coh}. Adler's theorem relates $\sigma_{COH}(\nu + X\rightarrow \nu + X + \pi^0)$ 
at $Q^2\rightarrow 0$ 
to $\sigma(\pi^0 + X \rightarrow \pi^0 + X)$. Subsequently, the model for the CC reaction, has been upgraded
\cite{23-coh_pcac_mass} 
to include lepton mass effects important for low $E_{\nu}$ studies. The new model predicts the $\sigma_{COH} (\pi^+)$/$\sigma_{COH} (\pi^0)$ 
ratio at $E_\nu=1$~GeV 
to be $1.45$ rather than $2$. Another important improvement was to use a better model for 
$d\sigma(\pi +\ {}^{12}C \rightarrow \pi +\ {}^{12}C)$/$dt$ in the region of pion kinematical energy 
$100$~MeV$<T_\pi<900$~MeV.  
As a result, the predicted COH cross section  from the model became reduced 
by a factor of 2-3 \cite{23-berger_sehgal}. The PCAC based approach 
was also discussed in \cite{23-coh_paschos} and critically re-derived in Ref.~\cite{23-Hernandez:2009vm}.

At lower energies the microscopic $\Delta$ dominance models for the COH reaction 
\cite{23-coh_micro}  are believed to be more reliable.
Within microscopic models there are still various approaches 
e.g due to differences in 
the treatment of the nonresonant background. The absolute normalization of the predicted cross-section depends 
on the adopted value of 
the $N\rightarrow \Delta$ form factor $C^A_5(0)$ because $\sigma_{COH}\sim (C^A_5(0))^2$.
\subsection{MC generators}
Almost all MC events generators rely on the old 
Rein-Sehgal resonance model for pion resonance production \cite{23-rs_res}. 
The model is based on the quark resonance model and includes 
contributions from 18 resonances covering the region $W<2$~GeV. 
The model is easily implementable in MC generators and it has only one set of vector and
axial form factors. 
In the original model, the charged 
lepton is assumed 
to be massless and prescriptions to cope with this problem 
were proposed in Refs. \cite{23-coh_pcac_mass,23-rs_lepton}. 
It was also realized that 
the RS model can be improved in the $\Delta$ region
by modyfying both vector and axial form factors using either old deuterium or new 
MiniBooNE pion production data \cite{23-jarek,23-rs_ff} . 

As for coherent pion production, all the MCs use the Rein-Sehgal COH model~\cite{23-rs_coh}. The analysis of 
of MC event generators and theoretical models done in \cite{23-Boyd:2009zz} show 
that in the $1-2$~GeV energy region, the Rein Sehgal COH model predictions disagree significantly
with all the recent theoretical computations and experimental results.

A crucial element of MC is the FSI model. These are typically semiclassical intra-nuclear cascade models. 
The topic of FSI goes far beyond the scope of this review and we only note that the progress in 
understanding the experimental data requires more reliable FSI models. The existing models should be 
systematically benchmarked with electro and photoproduction data as it was done in the case of GIBUU.
\subsection{Duality}
Bridging the region between RES and DIS (where with a good approximation interactions occur on quarks)
dynamics is a practical problem which must be resolved in all
MC event generators.  In MC event generators ``DIS'' is defined as ``anything but QE and RES'', what
is usually expressed as a condition on the invariant hadronic mass of the type $W>1.6$~GeV or $W>2$~GeV or so. 
Notice however that such a definition of ``DIS'' contains a contribution from the
kinematical region $Q^2<1$~GeV$^2$ which is beyond the applicability
of the genuine DIS formalism.  
RES/DIS transition region is not only a 
a matter of an arbitrary choice but is closely 
connected with the hypothesis of quark-hadron duality. 

Investigation of structure functions introduced in the formalism of the
inclusive electron-nucleon 
scattering led Bloom and Gilman to the observation that the average over resonances 
is approximately equal to the leading twist contribution measured in the completely different DIS region. One can
distinguish two aspects of duality: (i) resonant structure functions oscillate around a DIS scaling 
curve; (ii) the resonant structure functions for varying values of $Q^2$ 
{\it slide} along the DIS curve evaluated at fixed $Q^2_{DIS}$.

In order to quantify the degree in which the duality is satisfied one defines the ratio of integrals over
structure functions from RES and DIS:
\begin{equation}
{\cal R} (Q^2, Q^2_{DIS}) = \frac{\int_{\xi_{min}}^{\xi_{max}} d\xi F_j(\xi, Q^2)}
{\int_{\xi_{min}}^{\xi_{max}} d\xi F_j(\xi, Q^2_{DIS})}
\label{dual_ratio}
\end{equation}
The integrals are in the Nachtmann variable $\xi(x, Q^2) = \frac{2x}{1+\sqrt{1+4x^2M^2/Q^2}}$ 
and the integration region is defined as
$W_{min}<W<W_{max}$. Typically $W_{min}=M+m_\pi$ and $W_{max} = 1.6, ..., 2.0$~GeV. In the case of DIS,
the value of $Q^2_{DIS}$ is much larger and as a consequence the integral over $\xi$ runs over a quite different
region in $W$.

Neutrino-nucleon scattering duality studies are theoretical in their nature 
because the precise data in the resonance region are still missing. The duality was studied in three
papers: \cite{23-satolee1,23-olga_duality,23-wroclaw_duality}.
For neutrino interactions the duality can be satisfied only for the isospin average target. This is because the RES 
structure functions for proton are much larger than for neutron and in the case of DIS structure functions the
situation is opposite. 

Theoretical studies were done with a model which contains resonances from the first 
and second resonance region but not the background contribution and with the Rein-Sehgal model which
is commonly used in MC event generators. If the resonance region is confined to $W<1.6$~GeV the duality
as defined in Eq.~(\ref{dual_ratio}) is satisfied at the 75-80\% level. If the resonance region is extended to $W<2$~GeV
the value of the integral in Eq.~(\ref{dual_ratio}) is only about 50\%. These results are to some extent model dependent
but a general tendency is that for larger W, DIS structure functions are much larger than the resonance contribution, 
as clearly seen from Fig 3 in   \cite{23-olga_duality}   and Fig. 7 in \cite{23-wroclaw_duality}.
As shown in \cite{23-wroclaw_duality} 
there is also a 5\% uncertainty coming from an arbitrary choice of $Q^2_{DIS}$.

Two component duality hypothesis states that resonance contribution is dual to the valence quarks and the
nonresonant background to the sea. Investigation done within the Rein Sehgal model with $W<2$~GeV
revealed no signature of two component duality. Quark-hadron duality was also investigated in 
the case of neutrino nucleus interactions 
\cite{23-duality_nucleus}.

As a practical procedure for addressing this region, Bodek and Yang~\cite{B-Y}  have introduced and refined a model that is used by many contemporary neutrino event generators such as NEUGEN and its successor GENIE to bridge the kinematic region between the Delta and full DIS.   The model has been developed for both neutrino- and electron-nucleon inelastic scattering cross sections using leading order parton distribution functions and introducing a new scaling variable they call $\xi_w$. 

Non-perturbative effects that are prevalent in the kinematic region bridging the resonance and DIS regimes are described using the  $\xi_w$ scaling variable, in combination with multiplicative $K$ factors at low $Q^2$.  The model is successful in describing inelastic charged lepton-nucleon scattering,
including resonance production, from high-to-low $Q^2$.  In particular, the model describes existing
inelastic  neutrino-nucleon scattering measurements.   
 
Their proposed scaling variable, $\xi_w$ is derived using energy momentum conservation and assumptions about the initial/final quark mass and $P_T$.  Parameters are built into the derivation of  $\xi_w$ to account (on average)  for the higher order QCD terms and dynamic higher twist that is covered by an enhanced target mass term.

At the juncture with the DIS region, the Bodek-Yang model incorporates the GRV98~\cite{GRV98} LO parton distribution functions replacing the variable x with $\xi_w$.  They introduce "K-factors", different for sea and valence quarks, to multiply the PDFs so that they are correct at the low $Q^2$ photo-production limit.  A possible criticism of the model is the requirement of using the rather dated GRV98 parton distribution functions in the DIS region so the bridge to the lower W kinematic region is seamless.
\section{$\nu$-A Deep-inelastic Scattering: Introduction}
Although deep-inelastic scattering (DIS) is normally considered to be a topic for much higher energy neutrinos, 
wide-band beams such as the Fermilab NuMI and the planned LBNE beams do have real contributions from DIS that are 
particularly important in feed-down to the background that must be carefully considered.  In addition, there are 
x-dependent nuclear effects that should be taken into account when comparing results from detectors with different nuclei and even when comparing results from "identical" near and far detectors when the neutrino spectra entering the near and far detectors are different.

For this review, the definition of deep-inelastic scattering (DIS) is the kinematic based definition with 
W $\geq$ 2.0 GeV and $Q^2 \geq$ 1.0 GeV.  This is mostly out of the resonance production region and allows a fit to parton distribution functions. 
As said in Introduction, 
this is unfortunately not the definition used by several modern Monte Carlo generators that do not 
differentiate between simply "inelastic" interactions and deep-inelastic interactions calling everything beyond the Delta simply DIS.  This is an unfortunate confusing use of nomenclature by the generators.

In general, deep-inelastic scattering offers an opportunity to probe the partonic structure of the nucleon both in its free state and when the nucleon is bound in a nucleus.  Description of the partonic structure can include {\em parton distribution functions (PDFs)} giving the longitudinal, transverse and spin distributions of quarks within the nucleon  as well as, for example, the hadron formation zone giving the time/length it takes for a struck quark to fully hadronize into a strong-interacting hadron.

Neutrino scattering can play an important role in extraction of these fundamental parton distribution functions (PDFs) since only neutrinos via the weak-interaction can resolve the flavor of the nucleon's constituents: $\nu$ interacts with $d$, $s$, $\ubar$ and $\cbar$ while the $\overline{\nu}$ interacts with $u$, $c$, $\dbar$ and $\sbar$.  The weak current's unique ability to "taste" only particular quark flavors significantly enhances the study of parton distribution functions.  High-statistics measurement of the nucleon's partonic structure, using neutrinos, could complement studies with electromagnetic probes.

In the pursuit of precision measurements of neutrino oscillation parameters, large data samples and dedicated effort to minimize systematic errors could allow neutrino experiments to independently isolate all {\em SIX} of the weak structure functions $F_1^{\nu N}(x,Q^2)$, $F_1^{\bar \nu N}(x,Q^2)$, $F_2^{\nu  N}(x,Q^2)$, $F_2^{\bar \nu N}(x,Q^2)$, $xF_3^{\nu N}(x,Q^2)$ and $xF_3^{\bar \nu N}(x,Q^2)$ for the first time. 
Then, by taking differences and sums of these structure functions,  specific parton distribution functions in a given $(x, Q^2)$ bin can in turn be better isolated.  Extracting this full set of structure functions will rely on the $y$-variation of the structure function coefficients in the expression for the cross-section.
In the helicity representation, for example:
\begin{eqnarray}
\frac{d^2 \sigma^\nu}{dx d\qsq} &=& \frac{G^2_F}{2 \pi x}
\Bigl[\frac{1}{2}
\left( F_2^\nu (x,\qsq) + xF_3^\nu (x,\qsq) \right) + \nonumber \\
& & \frac{(1-y)^2}{2}
\left(F_2^\nu (x,\qsq) - xF_3^\nu (x,\qsq) \right) - 
2 y^2 F_L^\nu (x,\qsq) \Bigr].
\label{eq:pnp1}
\end{eqnarray}
where $F_L$ is the longitudinal structure function representing the absorption of longitudinally polarized Intermediate Vector Boson.

\noindent By analyzing the data as a function of $(1-y)^2$ in a given $(x, \qsq)$ bin for both $\nu$ and $\overline {\nu}$, all six structure functions could be extracted.

Somewhat less demanding in statistics and control of systematics, the ``average" structure functions $F_2(x,Q^2)$ and $xF_3(x,Q^2)$ can be determined from fits to combinations of the neutrino and antineutrino differential cross sections and several assumptions.  The sum of the $\nu$ and $\overline {\nu}$ differential cross sections, yielding $F_2$ then can be expressed as:
\begin{eqnarray}
{\frac{d^2\sigma}{dx dy}}^{\nu}+{\frac{d^2\sigma}{dx dy}}^{\overline\nu}
={\frac{G_{F}^2
M E}{\pi}}\Big[2\Big(1-y-{\frac{M x y}{2E}} 
 +  {\frac{y^2}{2}{\frac{1+4 M^2 x^2/ Q^2} {1+R_{L}}}}\Big) F_{2}  +  {y} \Big(1 - {\frac{y}{2}}\Big)  \Delta xF_{3}\Big]
\end{eqnarray}
where $R_L$ is equal to $\sigma_L $ / $\sigma_T$ and now $F_{2}$ is the {\em average} of $F_{2}^{\nu}$ and $F_{2}^{\overline {\nu}}$ and the last term is 
proportional to the difference in $xF_3$ for neutrino and antineutrino probes, 
$\Delta xF_3=xF_3^{\nu}-xF_3^{\overline{\nu}}$.  In terms of the strange and charm parton distribution function s and c,  at leading order, assuming symmetric $s$ and $c$ seas, this is 
$4x\left(s-c\right)$.  

The cross sections are also corrected for the excess of neutrons over protons in the target (for example the Fe correction is 5.67\%) so that the presented structure functions are for an isoscalar target.   
A significant step in the determination of $F_2(x,Q^2)$ in this manner that affects the low-x values is the assumed $\Delta xF_3$ and $R_L(x,Q^2)$.
Recent analyses use, for example, a NLO QCD model as input (TRVFS \cite{23-trvfs}) and assumes an input value of $R_L(x,Q^2)$ that comes from a fit to the world's charged-lepton measurements \cite{23-rworld}.  This could be an additional problem since, as will be suggested, $R_L(x,Q^2)$ can be different for neutrino as opposed to charged-lepton scattering. 

The structure function  $xF_3$ can be determined in a similar manner by taking the difference in $\nu$ and $\overline {\nu}$ differential cross sections.
\subsection{The Physics of Deep-inlastic Scattering}
There have been very few recent developments in the theory of deep-inelastic scattering.  The theory has been well-established for years.  The most recent developments in neutrino DIS scattering involve 
the experimental determination of parton distribution functions of nucleons within a nucleus, so-called {\em nuclear} parton distribution functions (nPDF).   
The more contemporary study of  $\nu$ nucleus deep-inelastic scattering using high-statistics experimental 
results with careful attention to multiple systematic errors began with the CDHSW, CCFR/NuTeV $\nu$ Fe, the NOMAD $\nu$ C and the CHORUS $\nu$ Pb experiments.  
Whereas NuTeV \cite{23-nutev} and CHORUS \cite{23-chorus} Collaborations have published their full data sets, NOMAD \cite{23-Altegoer:1997gv} has not yet done so.  
This short summary of DIS physics will concentrate on nuclear/nucleon parton distribution functions.
\subsubsection{Low-and-High $Q^2$ Structure Functions: Longitudinal and Transverse}
Since the current and future neutrino beams designed for neutrino oscillation experiments will be concentrating on lower energy neutrinos (1 - 5 GeV), 
many of the interactions will be at the lower-Q edge of DIS or even in the "soft" DIS region - namely, 
W $\geq$ 2.0 GeV however, with $Q^2 \leq 1.0$ GeV$^2$.   Understanding the physics of this kinematic region is 
therefore important.  

Since both the vector and axial-vector part of the transverse structure function $F_T$ go to 0 at $Q^2$ = 0 
(similar to $\ell^{\pm }$ charged-lepton vector current scattering), the low-$Q^2$ region $\nu$ and  $\overline \nu$ cross sections are dominated by the longitudinal structure function $F_L$.  The longitudinal structure function is composed of a vector and axial-vector component $F_L^{VC}$ and $F_L^{AC}$ and the low-$Q^2$ behavior of these components is not the same as in the transverse case.  The conservation of the vector current (CVC) suggests that  $F_L^{VC}$ behaves as the vector current in charged-lepton scattering and vanishes at low $Q^2$.  
However, the axial-vector current is not conserved and is related to the pion field via PCAC, so there is a surviving low $Q^2$ contribution from this component~\cite{Petti:2006tu} and $F_L^{AC}$ dominates the low $Q^2$ behavior.  Consequently, the ratio $R = F_L/F_T$ is divergent for neutrino interactions. This is substantially different from the scattering of charged leptons for which $R$ is vanishing as $Q^2$ and using measurement of $R$ from charged lepton scattering to determine $F_2$ for neutrino scattering is obviously wrong for lower Q.  In addition, this non-vanishing and dominant longitudinal structure function could be important for the interpretation of low-$Q^2$ nuclear effects with neutrinos to be described shortly.  
\subsubsection{Low-and-High $Q^2$ Structure Functions: 1/$Q^2$ Corrections}
Using a notation similar to that of reference~\cite{23-Accardi:2009br}, the total structure function can be expressed in a phenomenological form:
\begin{eqnarray}
F_i(x,Q^2)
&=& F_i^{\rm LT}(x,Q^2)
    \left( 1 + \frac{C_4(x)}{Q^2} \right)\, ,
\label{eq:HT}
\end{eqnarray}
where $i=1,2,3$ refers to the type of the structure function.  Using i = 2 as an example, then  $F_2^{\rm LT}$ is the leading twist component that has already included target mass corrections (TMC) and $C_4$ is the coefficient of the twist-4 term, the first higher-twist term proportional to $1/Q^2$).  There are, of course, further higher-twist terms $\frac{H_i^{(T=6)}(x)}{Q^4} + \cdots $ 
proportional to ever increasing powers of $1/Q^2$ however, for most phenomenological fits, the dominant leading twist plus twist-4 term are sufficient to describe the data.  The target mass corrections are kinematic in origin and involve terms suppressed by powers of $M^2/Q^2$ while the higher twist terms are dynamical in origin and are suppressed as mentioned by powers of $1/Q^2$.  These higher-twist terms are associated with multi-quark or quark and gluon fields and it is difficult to evaluate their magnitude and shape from first principles.    As with the kinematic target mass corrections, these must be taken into account in analyses of data at low $Q^2$ and especially at large $x$.  At higher $Q^2$ the contribution of the HT terms is negligible and there are various 
global fits~\cite{23-Pumplin:2002vw,23-Stirling:2005td} to the structure functions (among various scattering input) to determine the parton distribution functions (PDFs) that do not include any HT terms.  
 
The analysis of nuclear PDFs to be described shortly uses data from a TeVatron neutrino experiment at very high neutrino energies and thus is one of the analyses that does not need to be concerned with higher-twist corrections.  However, the current neutrino-oscillation oriented beam-lines are not high-energy and the analyses of this data may indeed need to consider both target mass corrections and higher-twist.  If indeed inclusion of higher-twist in these analyses becomes necessary, 
the authors of~\cite{23-Accardi:2009br} stress the importance of explicitly including {\em both} the target mass corrections {\it and} the higher twist corrections, even though they have very different physical origin and can have very different $x$ dependence.    It is important to note, as mentioned, that there are both {\em nucleon} and {\em nuclear} PDFs depending on the target.  The relation between them, called nuclear correction factors, are currently being studied for both  $\nu$-A and $\ell^{\pm }A$.  There are early indications that the nuclear correction factors for these two processes may not be the same.
\section{Recent DIS measurements: Neutrino Iron Scattering Results}
\begin{figure}[tbh]
\begin{center}
\includegraphics[width=0.5\textwidth]{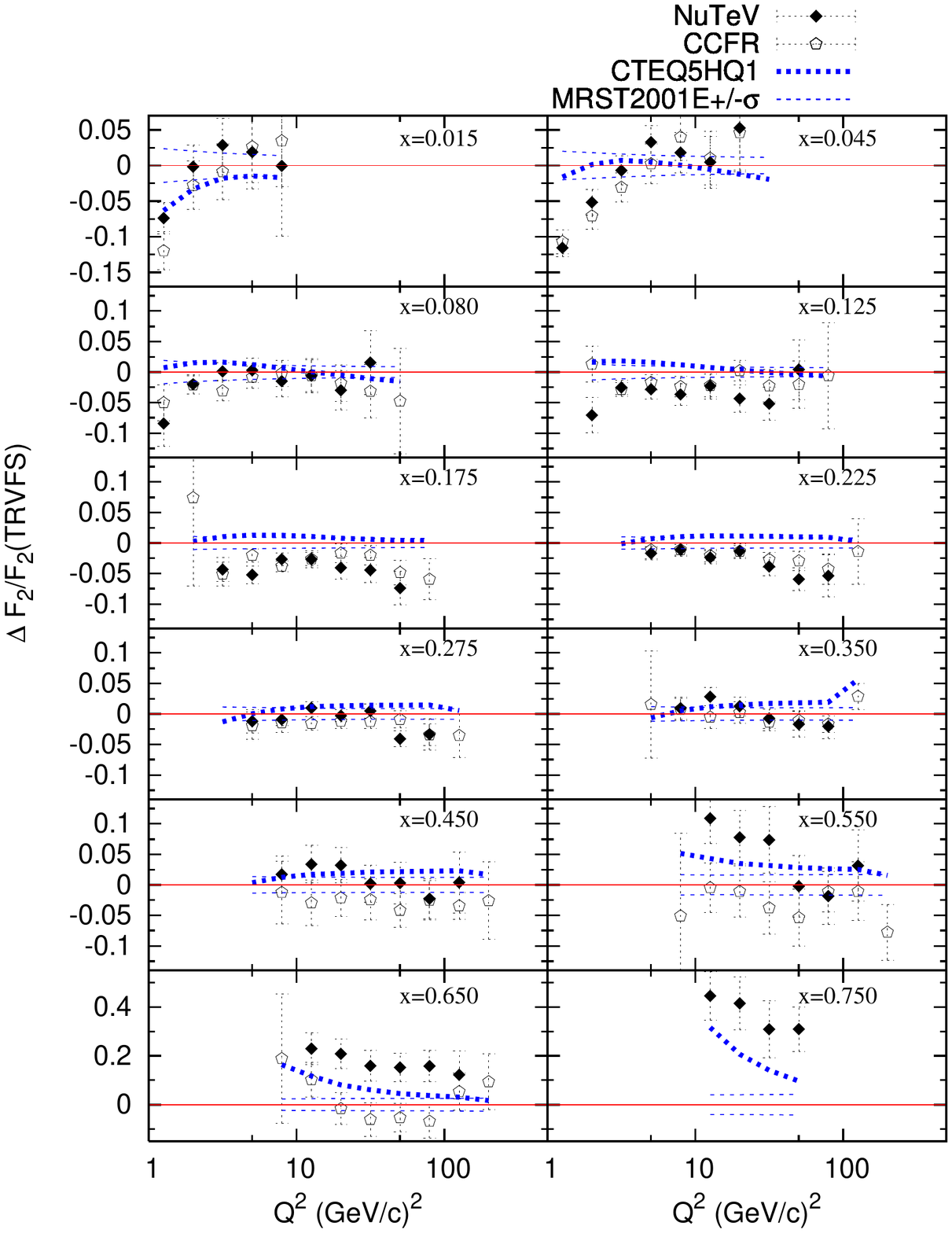}
\caption{ 
A comparison of the measurements of the $F_2$ structure function by NuTeV and CCFR and the predictions from the global PDF fits of the MRST and CTEQ collaboration~\cite{Tzanov:2005fu} that does not use the NuTeV data points as input to their fit.  The model predictions have already been corrected for target mass and, most significantly, nuclear effects {\em assuming these corrections are the same for charge-lepton and neutrino interactions}}
\label{fig-23:NuTeV-PDFs}
\end{center}
\end{figure}

The difficulty, of course, is that modern neutrino oscillation experiments demand high statistics which means that the neutrinos need massive nuclear targets to acquire these statistics.  This, in turn, complicates the extraction of free nucleon PDFs and demands nuclear correction factors that scale the results on a massive target to the corresponding result on a nucleon target. 
The results of the latest study of QCD using neutrino scattering comes from the NuTeV experiment \cite{23-nutev}. 
The NuTeV experiment was a direct follow-up of the CCFR experiment using nearly the same detector as CCFR but with a different neutrino beam.   
The NuTeV experiment accumulated over 3 million  $\nu$  and $\nub$ events in the energy range of 20 to 400 GeV off a manly Fe target. 
A comparison of the NuTeV results with those of CCFR and the predictions of the major PDF-fitting collaborations 
(CTEQ and MRST~\cite{23-Pumplin:2002vw,23-Stirling:2005td} ) is shown in Figure~\ref{fig-23:NuTeV-PDFs}.  

The main points are that the NuTeV $F_2$ agrees with CCFR  for values of $x_{B j} \leq$  0.4 but is systematically higher for larger values of $x_{B j}$ culminating at $x_{B j}$ = 0.65 where the NuTeV result is 20\% higher than the CCFR result.  NuTeV agrees with charged lepton data for $x_{B j} \leq$ 0.5 but there is increasing disagreement for higher values.
Although NuTeV $F_2$ and $xF_3$ agree with theory for medium x, they find a different $Q^2$ behavior at small x and are systematically higher than theory at high x.
These results can be summarized in four main questions to ask subsequent neutrino experiments:
\begin{itemize}
\item At high x, what is the behavior of the valence quarks as x  $\rightarrow$ 1.0?
\item At all x and $Q^2$, what is yet to be learned if we can measure all six $\nu$  and $\nub$ structure functions to yield maximal information on the parton distribution functions?
\item At all x, how do nuclear effects with incoming neutrinos differ from nuclear effects with incoming charged leptons?
\end{itemize}

This last item highlights an overriding question when trying to get a global view of structure functions from both neutrino and charged-lepton scattering data.  How do we compare data off nuclear targets with data off nucleons and, the associated question, how do we scale nuclear target data to the comparable nucleon data.
In most PDF analyses, the nuclear correction factors were taken from $\ell^\pm$-nucleus scattering and 
used for both charged-lepton and neutrino scattering.  
Recent studies by a CTEQ-Grenoble-Karlsruhe 
collaboration (called nCTEQ)~\cite{23-arXiv:1012.0286}
have shown that there may indeed be a difference between the charged-lepton and neutrino correction factors.

The data from the high-statistics $\nu$-DIS experiment, NuTeV summarized above, was used to perform a dedicated 
PDF fit to neutrino--iron data~\cite{23-npdf}.
The methodology for this fit is parallel to that of the previous global analysis 
\cite{23-owens} {\it but} with the difference that only Fe data has been used and no nuclear 
corrections have been applied to the analyzed data; hence, the resulting PDFs are for a proton in an iron 
nucleus - nuclear parton distribution functions\footnote{For more details of the fitting techniques and resulting comparisons with charged-lepton scattering see Part II of reference~\cite{Kopeliovich:2012fu}.}

\begin{figure}[tbh]
\begin{center}
\includegraphics[width=1.0\textwidth]{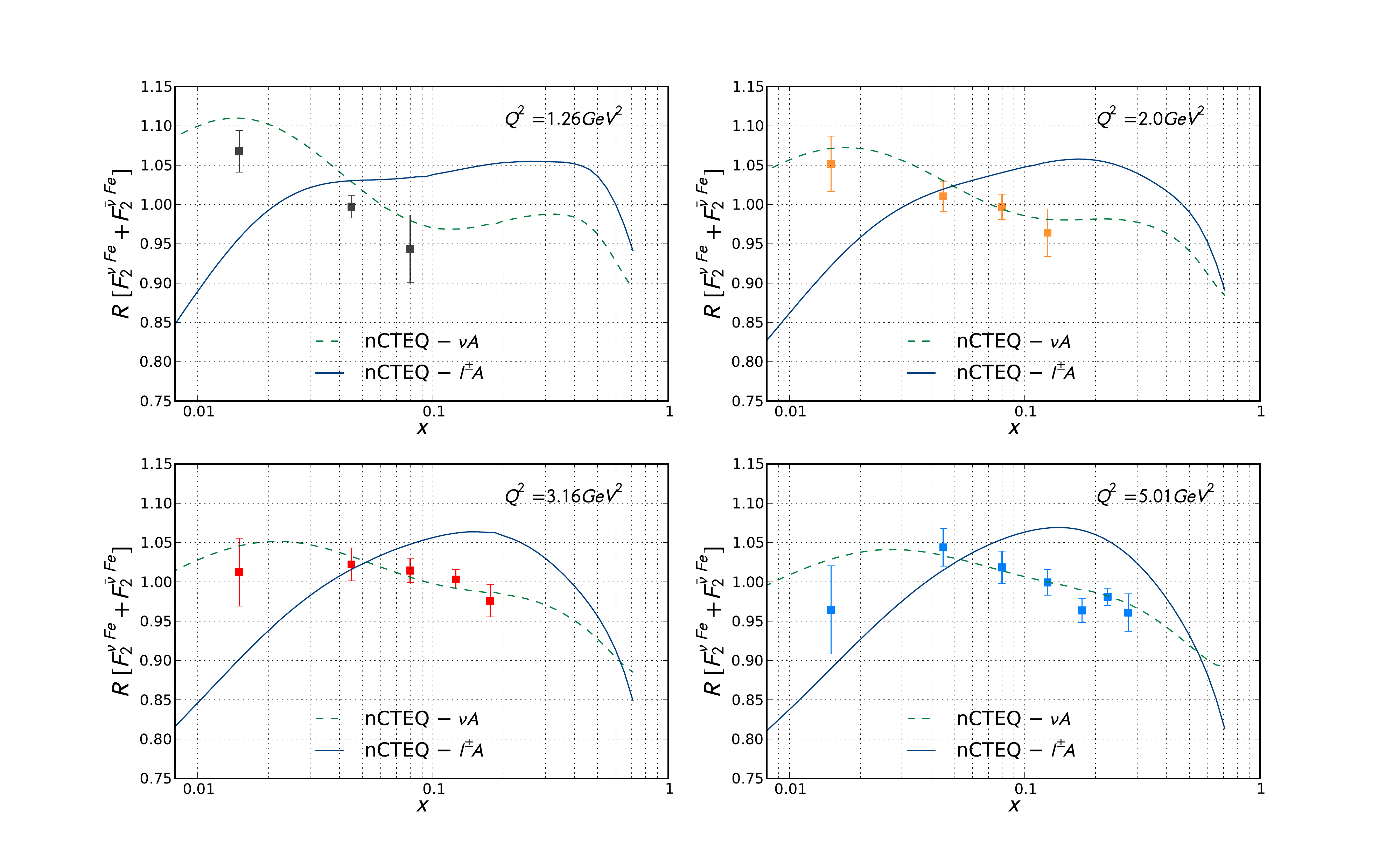}
\caption{ 
Nuclear correction factor $R$ for the average $F_2$ structure function in charged current $\nu Fe$ scattering at  $Q^2$ =1.2, 2.0, 3.2 and 5.0 $GeV^2$ compared to the measured NuTeV points.  The green dashed curve curve shows the result of the nCTEQ analysis of $\nu$ A (CHORUS, CCFR and NuTeV) differential cross sections plotted in terms of the average $F_2^{Fe}$ divided by the results obtained with the reference fit (free-proton) PDFs.  For comparison, the nCTEQ fit to the charged-lepton data is shown by the solid blue curve.}
\label{fig-23:F2lowQ}
\end{center}
\end{figure}

By comparing these iron PDFs with the free-proton PDFs (appropriately scaled) a neutrino-specific heavy target nuclear correction factor R can be obtained which should be applied to relate these two quantities.
It is also of course possible to combine these fitted nPDFs to form the individual values of the average of $F_2(\nu A)$ and $F_2(\bar\nu A)$ for a given x, $Q^2$ to compare directly with the NuTeV published values of this quantity.  This was recently done and the nCTEQ preliminary results~\cite{23-DIS12} for low-$Q^2$ are shown in Figure~\ref{fig-23:F2lowQ}.
Although the neutrino fit has general features in common with the charged-lepton parameterization, 
the magnitude of the effects and the $x$-region where they apply are quite different.  
The present results are noticeably flatter than the charged-lepton curves, especially at low- and 
moderate-$x$ where the differences are significant.  The comparison between the nCTEQ fit, 
that passes through the NuTeV measured points, and the charged-lepton fit is very different in the lowest-x, 
lowest-$Q^2$ region and gradually approaches the charged-lepton fit with increasing $Q^2$.  
However, the slope of the fit approaching the shadowing region from higher x where the 
NuTeV measured points and the nCTEQ fit are consistently below the charged-lepton A fit, 
make it difficult to reach the degree of shadowing evidenced in charged-lepton nucleus scattering at even higher $Q^2$. 

The general trend is that the anti-shadowing region is shifted to smaller $x$ values, and any 
turn-over at low $x$ is minimal given the PDF uncertainties.  
More specifically, there is no indication of "shadowing" in the NuTeV neutrino results at low-$Q^{2}$.
In general, these plots suggest that the size of the nuclear
corrections extracted from the NuTeV data are smaller than those
obtained from charged lepton scattering. 
\subsection{Comparison of the $\ell^{\pm}A$ and $\nu A$ Nuclear Correction Factors}
For the nCTEQ analysis, the contrast between the charged-lepton ($\ell^{\pm}A$) case and the neutrino ($\nu A$) case is striking.  While the nCTEQ fit to charged-lepton and Drell-Yan data generally align with the other charged-lepton determinations, the neutrino results clearly yield different behavior as a function of $x$, particularly in the shadowing/anti-shadowing region.   In the $\overline\nu$ case, these differences are smaller but persist in the low-x shadowing region.  The nCTEQ collaboration emphasize that both the charged-lepton and neutrino results come directly from global fits to the data, there is no model involved.  They further suggest that this difference between the results in charged-lepton and neutrino DIS is reflective of the long-standing {}``tension'' between the light-target charged-lepton data and the heavy-target neutrino data in the historical global 
PDF fits \cite{23-Botts:1992yi,23-Lai:1994bb}. Their latest results suggest that the tension is not only between charged-lepton \emph{light-target} data and neutrino heavy-target data, but also between neutrino and charged-lepton \emph{heavy-target} data.  In other words a difference between charged-lepton ($\ell^{\pm}A$) and the neutrino ($\nu A$) when comparing the same A.

Concentrating on this interesting difference found by the nCTEQ group, if the nuclear corrections for 
the $\ell^{\pm}A$ and $\nu A$ processes are indeed different there are several far-reaching consequences.  
Considering this, the nCTEQ group has performed a unified 
global analysis~\cite{23-arXiv:1012.0286} of the $\ell^{\pm}A$, DY, and $\nu A$ data 
(accounting for appropriate systematic and statistical errors) to determine if it is possible to obtain a 
{}``compromise'' solution including both $\ell^{\pm}A$ and $\nu A$ data. Using a hypothesis-testing criterion 
based on the $\chi^{2}$ distribution that can be applied to both the total $\chi^{2}$ as well as to the $\chi^{2}$ of 
individual data sets, they found it was {\em not possible} to accommodate the data from $\nu A$ and $\ell^{\pm}A$ DIS 
by an acceptable combined fit.  

That is, when investigating the results in detail, the tension between the $\ell^{\pm}Fe$ and $\nu Fe$ data sets permits {\em no possible compromise fit} which adequately describes the neutrino DIS data along with the charged-lepton data and, consequently, $\ell^{\pm}Fe$ and $\nu Fe$ based on the NuTeV results, have different nuclear correction factors.

A compromise solution between $\nu A$ and $\ell^{\pm}A$ data can be found {\em only} if the full correlated systematic errors of the $\nu A$ data are not used and the statistical and all systematic errors are combined in quadrature thereby neglecting the information contained in the correlation matrix.  In other words the larger errors resulting from combining statistical and all systematic errors in quadrature reduces the discriminatory power of the fit such that the difference between $\nu A$ and $\ell^{\pm}A$ data are no longer evident.  This conclusion 
underscores the fundamental difference~\cite{23-arXiv:1012.0286} of the nCTEQ analysis with other contemporary analyses.  

On the other hand, a difference between $\nu A$ and $\ell^{\pm}A$ is not completely unexpected, particularly in the shadowing region, and has previously been discussed in the literature \cite{23-Brodsky:2004qa, 23-Qiu:2004qk}.  The charged-lepton processes occur (dominantly) via $\gamma$-exchange, while the neutrino-nucleon processes occur via $W^{\pm}$-exchange. The different nuclear corrections could simply be a consequence of the differing propagation of the hadronic fluctuations of the intermediate bosons (photon, $W$) through dense nuclear matter.  Furthermore, since the structure functions in neutrino DIS and charged lepton DIS are distinct observables with different parton model expressions, it is clear that the nuclear correction factors will not be exactly the same.  What is, however, unexpected is the degree to which the $R$ factors differ between the structure functions $F_{2}^{\nu Fe}$ and $F_{2}^{\ell Fe}$. In particular the lack of evidence for shadowing in neutrino scattering at low $Q^2$ down to $x\sim0.02$ is quite surprising.

Should subsequent experimental results confirm the rather substantial difference between charged-lepton and neutrino scattering in the shadowing region at low-$Q^2$ it is interesting to speculate on the possible cause of the difference.  A recent study of EMC, BCDMS and NMC data by a
 Hampton University - Jefferson Laboratory collaboration~\cite{23-Guzey}  suggests that anti-shadowing in charged-lepton nucleus scattering may be dominated by the longitudinal structure function $F_L$.  As a by-product of this study, their figures hint that shadowing in the data of EMC, BCDMS and NMC $\mu$ A scattering was being led by the transverse cross section with the longitudinal component crossing over into the shadowing region at lower x compared to the transverse.   

As summarized earlier, in the low-$Q^2$ region, the neutrino cross section is dominated by the longitudinal structure function $F_L$ via axial-current interactions since $F_T$ vanishes as $Q^2$ as $Q^2 \rightarrow$ 0 similar to the behavior of charged lepton scattering.  If the results of the NuTeV analysis are verified, one contribution to the different behavior of shadowing at low-$Q^2$ demonstrated by $\nu$ A and $\ell$ A, in addition to the different hadronic fluctuations in the two interactions, could be due to the different mix of longitudinal and transverse contributions to the cross section  of the two processes in this kinematic region.

Another hypothesis of what is causing the difference between neutrino and charged-lepton shadowing results comes from  V. Guzey ~\cite{23-Guzey-2012} who speculates that at low x, low-$Q^2$  the value of y is close to unity and the neutrino interactions primarily probe the down and strange quarks.  This is very different than the situation with charged-lepton scattering where the contribution from down and strange quarks are suppressed by a factor of 1/4 compared to the up and charm. Therefore, the discrepancy between the observed nuclear shadowing in lepton-nucleus total cross section at small $x$ and shadowing in total neutrino-nucleus cross section could be caused by the absence of nuclear shadowing in the nuclear strange quark distributions as extracted from the neutrino-nucleus data or even the poor knowledge of the strange-quark distribution in the free-nucleon that affects the neutrino-nucleus ratio more than the charged-lepton ratio.

\vspace{0.5cm}
{\bf Acknowledgments}
\vspace{0.5cm}

 This research was supported by DGI and FEDER funds, under contracts
   FIS2011-28853-C02-02, and the Spanish
  Consolider-Ingenio 2010 Programme CPAN (CSD2007-00042),  by Generalitat
  Valenciana under contract PROMETEO/2009/0090 and by the EU
  HadronPhysics2 project, grant agreement no. 227431.

JTS (on leave from Wroc\l aw University and partially supported by grants NN202 368439 and DWM/57/T2K/2007) thanks 
T. Golan for making Figure 1.

Fermilab is operated by Fermi Research Alliance, LLC under Contract No. De-AC02-07CH11359 with the United States Department of Energy.  JGM thanks Martin Tzanov, Louisiana State University, for permission to use Figure 6.


\begin{thebibliography}{1}


\bibitem{Gallagher:2011zz} 
  H.~Gallagher, G.~Garvey and G.~P.~Zeller,
  ``Neutrino-nucleus interactions,''
  Ann.\ Rev.\ Nucl.\ Part.\ Sci.\  {\bf 61}, 355 (2011).

\bibitem{23-inclusive}
Q. Wu, et al [NOMAD collaboration]
{\it ``A Precise Measurement of the Muon Neutrino-Nucleon Inclusive Charged Current
Cross-Section off an Isoscalar Target in the Energy Range $2.5<E_\nu<40$~GeV by NOMAD'',}
Phys. Lett. B{\bf 660} (2008) 19;
%
P. Adamson et al [MINOS collaboration],
 {\it ``Neutrino and antineutrino inclusive charged-current cross section measurements with the MINOS
near detector'',}
Phys. Rev. D{\bf 81} (2010) 072002; 
%
Y. Nakajima, et al [SciBooNE collaboration]
{\it ``Measurement of inclusive charged current interactions on carbon in a few-GeV neutrino beam'',}
Phys. Rev. D{\bf 83} (2011) 012005.
%
M. Ravonel Salzgeber [T2K collaboration], {\it ``Measurement of the flux averaged Inclusive Charged Current
cross section''}, poster presented at NEUTRINO2012 conference, Kyoto, June 4-9, 2012.

\bibitem{23-nondipole} B. Bhattacharya, R.J. Hill, and G. Paz, {\it ``Model independent determination of the axial 
mass parameter in quasielastic neutrino-nucleon scattering''}, Phys. Rev. D{\bf 84} (2011) 073006.

\bibitem{23-bodek_MA} A. Bodek, S. Avvakumov, R. Bradford, and H.S. Budd, 
{\it ``Extraction of the axial nucleon form-factor from neutrino experiments on deuterium''},
Eur. Phys. J. C{\bf 63} (2009) 355.

\bibitem{23-Bernard:2001rs} 
  V.~Bernard, L.~Elouadrhiri and U.~.G.~Meissner,
  {\it ``Axial structure of the nucleon: topical review'', }
  J.\ Phys.\ G G {\bf 28}, R1 (2002).

\bibitem{23-k2k_oxygen_MA} R. Gran, E.J. Jeon et al [K2K collaboration], {\it ``Measurement
of the quasielastic axial vector mass in neutrino interactions on oxygen''},
Phys. Rev. D {\bf 74} (2006) 052002.

\bibitem{23-k2k_carbon_MA} X. Espinal and F. Sanchez, {\it ``Measurement of the axial vector mass in neutrino-carbon interactions at K2K''},
AIP Conf. Proc. {\bf 967} (2007) 117.

\bibitem{23-minos_MA} M. Dorman [MINOS collaboration], {\it ``Preliminary results for CCQE scattering with the MINOS
near detector''}, AIP Conf. Proc. {\bf 1189} (2009) 133.

\bibitem{23-AguilarArevalo:2010zc}
  A.~A.~Aguilar-Arevalo et al.  [MiniBooNE Collaboration],
  {\it ``First Measurement of the Muon Neutrino Charged Current Quasielastic Double
  Differential Cross Section'', }
  Phys.\ Rev.\  D {\bf 81} (2010) 092005.

\bibitem{23-Lyubushkin:2008pe} 
  V.~Lyubushkin et al.  [NOMAD Collaboration],
  {\it ``A Study of quasi-elastic muon neutrino and antineutrino scattering in the NOMAD experiment'', }
  Eur.\ Phys.\ J.\ C {\bf 63}, 355 (2009).

\bibitem{23-minos_nuint11}  N. Mayer and N. Graf, {\it ``Improvements to MINOS CCQE Analysis''}, AIP Conf. Proc. {\bf 1405} (2011) 41.

\bibitem{23-sciboone_nuint11} Y. Nakajima [SCiBooNE collaboration], 
{\it ``Measurement of CC and CCQE Interactions at SciBooNE''}, AIP Conf. Proc. {\bf 1405} (2011) 47.

\bibitem{23-alcaraz_thesis} .L. Alcaraz Aunion {\it Measurement of the absolute 
$\nu_\mu$-CCQE cross section at the SciBooNE experiment} 
PhD Thesis supervised by F. S\'{a}nchez, Barcelona, July 2010, 
http://lss.fnal.gov/archive/thesis/fermilab-thesis-2010-45.pdf.

\bibitem{23-miniboone_nuint11} J. Grange [MiniBooNE collaboration], 
{\it ``New Results from MiniBooNE Charged Current Quasi-Elastic Anti-Neutrino Data''}, AIP Conf. Proc. {\bf 1405} (2011) 83.

\bibitem{23-minerva_ccqe} J.G. Morfin [MINERvA collaboration] {\it ``Inclusive neutrino cross section measurements at MINERvA''},
poster presented at NEUTRINO2012 conference, Kyoto, June 4-9, 2012.

\bibitem{23-nuance} D. Casper, {\it ``The Nuance neutrino physics simulations, and the future''},
Nucl. Phys. B (Proc. Suppl.) {\bf 112} (2002) 161;  

\bibitem{23-Oset:1987re}
  E.~Oset and L.~L.~Salcedo,
  {\it ``$\Delta$ selfenergy in nuclear matter'', }
  Nucl.\ Phys.\  A {\bf 468} (1987) 631.

\bibitem{23-jarek} J.A. Nowak, {\it ``Four Momentum Transfer Discrepancy in the Charged Current 
$\pi^+$ Production in the MiniBooNE: Data vs. Theory,''} AIP Conf.Proc.1189: 243 (2009).

\bibitem{23-Boyd:2009zz}
  S.~Boyd, S.~Dytman, E.~Hernandez, J.~Sobczyk and R.~Tacik,
  {\it ``Comparison of models of neutrino-nucleus interactions'', }
  AIP Conf.\ Proc.\  {\bf 1189} (2009) 60.

\bibitem{23-Smith:1972xh}
  R.~A.~Smith and E.~J.~Moniz,
  {\it ``Neutrino reactions on nuclear targets'', }
  Nucl.\ Phys.\ B {\bf 43} (1972) 605
   [Erratum-ibid.\ B {\bf 101} (1975) 547].

\bibitem{23-Benhar:1994hw}
  O.~Benhar, A.~Fabrocini, S.~Fantoni and I.~Sick,
  {\it ``Spectral function of finite nuclei and scattering of GeV electrons'', }
  Nucl.\ Phys.\ A {\bf 579} (1994) 493.

\bibitem{23-Benhar:2006nr}
  O.~Benhar and D.~Meloni,
  {\it ``Total neutrino and antineutrino nuclear cross-sections around 1-GeV'', }
  Nucl.\ Phys.\ A {\bf 789} (2007) 379.


\bibitem{23-Ankowski:2007uy}
  A.~M.~Ankowski and J.~T.~Sobczyk,
  {\it ``Construction of spectral functions for medium-mass nuclei'', }
  Phys.\ Rev.\ C {\bf 77} (2008) 044311.

\bibitem{23-Leitner:2008ue}
  T.~Leitner, O.~Buss, L.~Alvarez-Ruso and U.~Mosel,
  {\it ``Electron- and neutrino-nucleus scattering from the quasielastic to the resonance region'', }
  Phys.\ Rev.\ C {\bf 79} (2009) 034601.


\bibitem{23-Nieves:2004wx}
  J.~Nieves, J.~E.~Amaro and M.~Valverde,
  {\it ``Inclusive quasi-elastic neutrino reactions'', }
  Phys.\ Rev.\ C {\bf 70} (2004) 055503
   [Erratum-ibid.\ C {\bf 72} (2005) 019902].

\bibitem{23-Nieves:2005rq}
  J.~Nieves, M.~Valverde and M.~J.~Vicente Vacas,
  {\it ``Inclusive nucleon emission induced by quasi-elastic neutrino-nucleus interactions'', }
  Phys.\ Rev.\ C {\bf 73} (2006) 025504.


\bibitem{23-Martini:2009uj}
M.~Martini et al., {\it ``A Unified approach for nucleon knock-out, coherent and incoherent pion production in neutrino interactions with nuclei,''}, Phys. Rev. {\bf C80} (2009) 065501.

\bibitem{23-Martini:2010ex} M.~Martini, M.~Ericson, G.~Chanfray, and
J.~Marteau, {\it ``Neutrino and antineutrino quasielastic interactions with nuclei,''}, Phys. Rev. {\bf C81} (2010) 045502.

\bibitem{23-aligarh} M. Sajjad Athar, S. Chauhan, and S.K. Singh, {\it ``Theoretical study of 
lepton events in the atmospheric neutrino 
experiments at SuperK''}, Eur. Phys. J. A{\bf 43} (2010) 209.

\bibitem{23-beta} D.~H.~ Wilkinson, {\it ``Renormalization of the axial-vector
coupling constant in nuclear $\beta$-decay,'' }, Nucl. Phys. {\bf
A209} (1973) 470; {\bf A225} (1974) 365.

\bibitem{23-Nieves:2011yp}
  J.~Nieves, I.~Ruiz Simo and M.~J.~Vicente Vacas,
  {\it ``The nucleon axial mass and the MiniBooNE Quasielastic Neutrino-Nucleus Scattering problem'', }
  Phys.\ Lett.\ B {\bf 707} (2012) 72.

\bibitem{23-Martini:2011wp}
  M.~Martini, M.~Ericson and G.~Chanfray,
  {\it ``Neutrino quasielastic interaction and nuclear dynamics'', }
  Phys.\ Rev.\ C {\bf 84} (2011) 055502.


\bibitem{23-AguilarArevalo:2008yp}
  A.~A.~Aguilar-Arevalo  et al.  [MiniBooNE Collaboration],
  {\it ``The Neutrino Flux prediction at MiniBooNE'', }
  Phys.\ Rev.\ D {\bf 79} (2009) 072002,

\bibitem {23-crpa} N. Jachowicz, C. Praet, and J. Ryckebusch, {\it ``Modeling Neutrino-Nucleus 
Interactions in the few-GeV Regime''}, Acta Phys. Pol. B{\bf 40} (2009) 2559.


\bibitem{23-Jachowicz:2007ek}
  N.~Jachowicz, P.~Vancraeyveld, P.~Lava, C.~Praet and J.~Ryckebusch,
  {\it ``Strangeness content of the nucleon in quasielastic neutrino-nucleus reactions'', }
  Phys.\ Rev.\ C {\bf 76} (2007) 055501.

\bibitem{23-Maieron:2003df}
  C.~Maieron, M.~C.~Martinez, J.~A.~Caballero and J.~M.~Udias,
  {\it ``Nuclear model effects in charged current neutrino nucleus quasielastic scattering'', }
  Phys.\ Rev.\ C {\bf 68} (2003) 048501.


\bibitem{23-Amaro:2011qb}
  J.~E.~Amaro  et al.  {\it ``Relativistic analyses of quasielastic neutrino cross sections at MiniBooNE kinematics'', }
  Phys.\ Rev.\ D {\bf 84} (2011) 033004.

\bibitem{23-sw} B. Serot and J. Walecka,{\it ``The Realtivistic Many-Body Problem''},  Adv. Nucl. Phys. {\bf 16} (1986)
  1.

\bibitem{23-Meucci:2011ce}
  A.~Meucci, C.~Giusti and F.~D.~Pacati,
  {\it ``Relativistic descriptions of final-state interactions in neutral-current neutrino-nucleus scattering at MiniBooNE kinematics'', }
  Phys.\ Rev.\ D {\bf 84} (2011) 113003.

\bibitem{23-super}
T~.W~. Donnelly and I. Sick, {\it ``Superscaling in inclusive electron - nucleus scattering'',}
 Phys. Rev. Lett. {\bf 82} (1999) 3212;
Phys. Rev. C 60 (1999) 065502.

\bibitem{23-Amaro:2004bs}
  J.~E.~Amaro  et al.,
  {\it ``Using electron scattering superscaling to predict charge-changing neutrino cross sections in nuclei'', }
  Phys.\ Rev.\ C {\bf 71} (2005) 015501.

\bibitem{23-Amaro:2006pr}
  J.~E.~Amaro, M.~B.~Barbaro, J.~A.~Caballero and T.~W.~Donnelly,
  {\it ``Superscaling and neutral current quasielastic neutrino-nucleus scattering'', }
  Phys.\ Rev.\ C {\bf 73} (2006) 035503.


\bibitem{23-AlvarezRuso:2010ia}
  L.~Alvarez-Ruso,
  {\it ``Neutrino interactions: challenges in the current theoretical picture'', }
  arXiv:1012.3871 [nucl-th].



\bibitem{23-Benhar:2009wi}
  O.~Benhar and D.~Meloni,
  {\it ``Impact of nuclear effects on the determination of the nucleon axial mass'', }
  Phys.\ Rev.\  D {\bf 80}, 073003 (2009).


\bibitem{23-Benhar:2010nx}
  O.~Benhar, P.~Coletti and D.~Meloni,
  {\it ``Electroweak nuclear response in quasi-elastic regime'', }
  Phys.\ Rev.\ Lett.\  {\bf 105} (2010) 132301.



\bibitem{23-Juszczak:2010ve}
  C.~Juszczak, J.~T.~Sobczyk and J.~\.Z muda,
  {\it ``On extraction of value of axial mass from MiniBooNE neutrino quasi-elastic
  double differential cross section data'', }
  Phys.\ Rev.\  C {\bf 82} (2010) 045502.

\bibitem{23-Butkevich:2010cr}
  A.~V.~Butkevich,
  {\it ``Analysis of flux-integrated cross sections for quasi-elastic neutrino
  charged-current scattering off $^{12}$C at MiniBooNE energies'', }
  Phys.\ Rev.\  C {\bf 82} (2010) 055501.

\bibitem{23-marteau} J. Marteau, {\it ``Effects of the nuclear correlations on the neutrino oxygen interactions''}, Eur. Phys. J.
A{\bf 5} (1999) 183. 


\bibitem{23-Nieves:2011pp} 
  J.~Nieves, I.~Ruiz Simo and M.~J.~Vicente Vacas,
  {\it ``Inclusive Charged--Current Neutrino--Nucleus Reactions'', }
  Phys.\ Rev.\ C {\bf 83}, 045501 (2011).


\bibitem{23-Alberico:1983zg} 
  W.~M.~Alberico, M.~Ericson and A.~Molinari,
  {\it ``The Role Of Two Particles - Two Holes Excitations In The Spin - Isospin Nuclear Response'', }
  Annals Phys.\  {\bf 154}, 356 (1984).

\bibitem{23-Amaro:2010sd}
  J.~E.~Amaro et al.,
  {\it ``Meson-exchange currents and quasielastic neutrino cross sections in the SuperScaling Approximation model'', }
  Phys.\ Lett.\ B {\bf 696} (2011) 151.

\bibitem{23-Amaro:2011aa}
  J.~E.~Amaro, M.~B.~Barbaro, J.~A.~Caballero and T.~W.~Donnelly,
  {\it ``Meson-exchange currents and quasielastic antineutrino cross sections in the SuperScaling Approximation'', }
  Phys.\ Rev.\ Lett.\  {\bf 108} (2012) 152501.

\bibitem{23-bode_mec} A. Bodek, H.S. Budd, and M.E. Christie, {\it ``Neutrino Quasielastic Scattering on Nuclear Targets:
Parametrizing Transverse Enhancement (Meson Exchange Currents)''}, Eur. Phys. J. C{\bf 71} (2011) 1726.

\bibitem{23-Lalakulich:2012ac}
  O.~Lalakulich, K.~Gallmeister and U.~Mosel,
  {\it ``Many-Body Interactions of Neutrinos with Nuclei - Observables'', }
  Phys. \ Rev. \ C{\bf86} (2012) 014614

\bibitem{23-Sobczyk:2012ms}
  J.~T.~Sobczyk,
  {\it ``Multinucleon ejection model for Meson Exchange Current neutrino interactions'', }
  Phys. \ Rev. \  C{\bf86} (2012) 015504

\bibitem{23-pion_fsi_neut} P. de Perio, {\it ``NEUT pion FSI''}, AIP Conf. Proc. {\bf 1405} (2011) 223.

\bibitem{23-nuwro_fsi} T. Golan, C. Juszczak, and J.T. Sobczyk, 
{\it ``Final State Interactions Effects in Neutrino-Nucleus Interactions''}, 
 Phys. \ Rev.\  C{\bf86} (2012) 015505


\bibitem{23-Meloni:2012fq}
  D.~Meloni and M.~Martini,
  {\it ``Revisiting the T2K data using different models for the neutrino-nucleus cross sections'', }
  Phys. \ Lett. \ B{\bf716} (2012) 186

\bibitem{23-Martini:2012fa} 
  M.~Martini, M.~Ericson and G.~Chanfray,
  {\it ``Neutrino energy reconstruction problems and neutrino oscillations'', }
  Phys.\ Rev.\ D {\bf 85},  (2012) 093012.

\bibitem{23-Nieves:2012yz}
  J.~Nieves, F.~Sanchez, I.~R.~Simo and M.~J.~V.~Vacas,
  {\it ``Neutrino Energy Reconstruction and the Shape of the CCQE-like Total Cross Section'', }
 Phys.\ Rev.\ D {\bf 85},  (2012) 113008.

\bibitem{23-Mosel:2012hr} 
  U.~Mosel and O.~Lalakulich,
  {\it ``Neutrino-Long-Baseline Experiments and Nuclear Physics'', }
  arXiv:1204.2269 [nucl-th].

\bibitem{23-mb_elastic} A.A. Aguilar-Arevalo et al., [MinoBooNE collaboration] 
{\it ``Measurement of the Neutrino Neutral-Current 
Elastic Differential Cross Section''}, Phys. Rev. D{\bf 82}, 092005 (2010).

\bibitem{23-ncel_butkevich} A.V. Butkevich and D. Perevalov, {\it ``Neutrino neutral current elastic scattering on 12C''}, 
Phys. Rev. C{\bf 84} (2011) 015501.

\bibitem{23-ncel_benhar} O. Benhar and G. Veneziano, {\it ``Nuclear effects in neutral current quasi-elastic neutrino interactions''},
Phys. Lett. B{\bf 702} (2011) 326.

\bibitem{23-ncel_ankowski} A.M. Ankowski, {\it ``Consistent analysis of neutral- and charged-current neutrino scattering off carbon''},
arXiv:12-5:4804[nucl-th].

\bibitem{23-mb_coh} A.A. Aguilar-Arevalo et al., [MiniBooNE collaboration] 
{\it ``First Observation of Coherent $\pi^0$ Production 
in Neutrino Nucleus Interactions with $E_\nu<2$~GeV}, Phys. Lett. B{\bf 664}, 41 (2008).

\bibitem{23-k2k_ncpi0} S. Nakayama et al. [K2K collaboration], {\it ``Measurement of the single pi0 production in neutral current 
neutrino interactions with water by a 1.3 GeV wide band muon neutrino beam''}, Phys. Lett. B{\bf 619} 255.

\bibitem{23-mb_ncpi0} A.A. Aguilar-Arevalo et al [MiniBooNE collaboration], 
{\it ``Measurement of $\nu_\mu$ and $\bar\nu_\mu$ induced neutral current single $\pi^0$ production cross 
sections on mineral oil at $E_\nu\sim O(1 GeV)$''}
Phys. Rev. D{\bf 81} (2010)
     013005.

\bibitem{23-sb_ncpi0} Y. Kurimoto, et al [SciBooNE collaboration], {\it ``Measurement of 
Inclusive Neutral Current Neutral Pion Production on Carbon in a Few-GeV Neutrino Beam,''} 
Phys. Rev. D{\bf 81} (2009) 033004.


\bibitem{23-mb_ccpiplus} A.A. Aguilar-Arevalo et al., [MiniBooNE collaboration] 
{\it ``Measurement of Neutrino-Induced 
Charged-Current Charged Pion Productio, Phys. Lett. {\bf B250} (1990) 193.n Cross Sections on 
Mineral Oil at $E_\nu$~1 GeV''},  Phys. Rev. D{\bf 83} 052007 (2011).

\bibitem{23-mb_ccpi0} A.A. Aguilar-Arevalo et al., [MiniBooNE collaboration] {\it ``
Measurement of $\nu_\mu$ Induced Charged Current Neutral Pion Production Cross-Sections on 
Mineral Oil at $E_\nu\in 0.5-2.0$~GeV''}, Phys. Rev. D{\bf 83} 052009 (2011).

\bibitem{23-mb_ratio} A.A. Aguilar-Arevalo et al., [MiniBooNE collaboration] {\it ``
Measurement of the $\nu_\mu$ CC $\pi^+/QE$ Cross Section Ratio on 
Mineral Oil in a $0.8$~GeV Neutrino Beam''}, Phys. Rev. Lett. {\bf 103}, 081801 (2009).

\bibitem{23-k2k_ratio} A. Rodriguez, L. Whitehead, et al [K2K collaboration], {\it ``Measurement of single charged pion
production in the charged-current interactions of neutrinos in a 1.3 GeV wide band beam''},
Phys. Rev D{\bf 78} (2008) 032003.

\bibitem{23-gibuu_pions1} O. Lalakulich, K. Gallmeister, T. Leitner, and U. Mosel, 
{\it ``Pion production in the MiniBooNE''}, AIP Conf.Proc. 1405 (2011) 127.

\bibitem{23-leitner_luis_mosel} T. Leitner, O. Buss, L. Alvarez-Ruso, and U. Mosel,	
{\it ``Electron- and neutrino-nucleus scattering from the quasielastic to the resonance region''},
Phys.Rev. C{\bf 79} (2009) 034601.

\bibitem {23-gibuu_ratio} T. Leitner, O. Buss, U. Mosel, and L. Alvarez-Ruso, {\it ``Neutrino-induced pion production at
energies relevant for te MiniBooNE and K2K experiments''}, Phys. Rev. C{\bf 79} (2009) 038501.

\bibitem{athar_ratio} M. Sajjad Athar, S. Chauhan, and S.K. Singh, {\it ``CC1$1\pi^+$ to CCQE cross sections 
ratio at accelarator energies''}, J. Phys. G{\bf 37} (2010) 015005.

\bibitem{mashnik} M.J. Vicente Vacas, M.Kh. Khankhasaev, and S.G. Mashnik,	
{\it ``Inclusive pion double charge exchange above .5-GeV''}, nucl-th/9412023.

\bibitem{23-singh_coh} S.K. Singh, M. Sajjad Athar, and S. Ahmad, {\it ``Nuclear Effects in Neutrino Induced Coherent Pion Production
at K2K and MiniBooNE''},  Phys. Rev. Lett. {\bf 96} (2006) 241801.

\bibitem{23-wro_war} K.M. Graczyk, D. Kie\l czewska, P. Przew\l ocki, J.T. Sobczyk,
{\it ``$C^5_A$ axial form factor from bubble chamber experiments''},
Phys. Rev. D{\bf 80} (2009) 093001.

\bibitem{23-nieves_chft} E. Hernandez, J. Nieves, and M. Valverde, 
{\it ``Weak Pion Production off the Nucleon''}, 
Phys. Rev. D{\bf 76} (2007) 033005.

\bibitem{23-olga1} O. Lalakulich, and E.A. Paschos,
{\it ``Resonance production by neutrinos. I. J = 3/2 resonances''}, Phys.Rev. D{\bf 71} (2005) 074003.

\bibitem{23-olga_anl_bnl} O. Lalakulich, T. Leitner, O. Buss, and U. Mosel, {\it ``One pion production
in neutrino reactions: Including nonresonant background''}, Phys. Rev. D{\bf 82} (2010) 093001.

\bibitem{23-singh_deu} L. Alvarez-Ruso, S.K. Singh, and M.J. Vicente Vacas,
{\it ``Neutrino d  $\rightarrow \mu^- \Delta^{++}$ n reaction and axial vector $N\Delta$ coupling''},
Phys. Rev. C{\bf 59} (1999) 3386.

\bibitem{23-goodness} M. Maltoni and T. Schwetz, 	
{\it ``Testing the statistical compatibility of independent data sets''}, Phys. Rev. D{\bf 68} (2003)
033020.

\bibitem{23-nieves_deut} E. Hernandez, J. Nieves, M. Valverde, and M.J. Vicente Vacas,	
{\it ``$N-\Delta(1232)$ axial form factors from weak pion production''},
Phys. Rev. D{\bf 81} (2010) 085046.

\bibitem{23-satolee1} T. Sato, D. Uno, and T.-S.H. Lee,
{\it ``Dynamical model of weak pion production reactions''}, Phys. Rev. C{\bf 67} (2003) 065201.

\bibitem{23-barbero} C. Barbero, G. Lopez Castro, and A. Mariano,
{\it ``Single pion production in CC $\nu_\mu N$ scattering within a consistent effective Born approximation''},
Phys. Lett. B{\bf 664} (2008) 70.

\bibitem{23-sb_coh_nc} SciBooNE collaboration, {\it ``Improved Measurement of Neutral Current Coherent 
$\pi^0$ Production on Carbon in a Few-GeV Neutrino Beam,''} Phys. Rev. D{\bf 81}, 111102(R) (2010).

\bibitem{23-k2k_coh} M. Hasegawa et al [K2K collaboration], {\it ``
Search for coherent charged pion production in neutrino-carbon interactions''},
Phys. Rev. Lett. 95, 252301 (2005). 

\bibitem{23-sb_coh_cc} SciBooNE collaboration, {\it ``Search for Charged Current Coherent 
Pion Production on Carbon in a Few-GeV Neutrino Beam,''} Phys. Rev. D{\bf 78}, 112004 (2008).

\bibitem{23-rs_coh} D. Rein and L.M. Sehgal, {\it ``Coherent pi0 Production in Neutrino Reactions''}, 
Nucl. Phys. B{\bf 223} (1983) 29.

\bibitem{23-coh_pcac_mass} 
Ch. Berger, L.M. Sehgal, {\it ``Lepton mass effects in single pion production by neutrinos''},
Phys. Rev. D{\bf 76} (2007) 113004.

\bibitem{23-berger_sehgal} 
Ch. Berger, L.M. Sehgal, {\it ``PCAC and coherent pion production by low energy neutrinos,''}
Phys. Rev. D{\bf 79} (2009) 053003.

\bibitem{23-coh_paschos} E.A. Paschos, and D. Schalla, {\it ``Coherent Pion Production by Neutrinos''},
Phys. Rev. D{\bf 80} (2009) 033005.

\bibitem{23-Hernandez:2009vm} E.Hernandez, J.Nieves and M.J.Vicente-Vacas,
  {\it ``Neutrino Induced Coherent Pion Production off Nuclei and PCAC,''}
  Phys. Rev. D {\bf 80} (2009) 013003.

\bibitem{23-coh_micro} 
S.X. Nakamura, T. Sato, T.-S.H. Lee, B. Szczerbinska, and K. Kubodera,
{\it ``Dynamical Model of Coherent Pion Production in Neutrino-Nucleus Scattering''},
Phys. Rev. C{\bf 81} (2010) 035502; J.E.Amaro, E.Hernandez, J.Nieves and M.Valverde,
  {\it `Theoretical study of neutrino-induced coherent pion production off nuclei at T2K and MiniBooNE energies``},
  Phys. Rev. D {\bf 79} (2009) 013002; 
E. Hernandez, J. Nieves, and M. Valverde, {\it ``
Coherent pion production off nuclei at T2K and MiniBooNE energies revisited''},
Phys.Rev. D{\bf 82} (2010) 077303; 
L. Alvarez-Ruso, L.S. Geng, S. Hirenzaki, and M.J. Vicente Vacas, {\it ``
Charged current neutrino induced coherent pion production''}, 
Phys.Rev. C{\bf 75} (2007) 055501, Erratum-ibid. C{\bf 80} (2009) 019906.

\bibitem{23-rs_res} D. Rein and L.M. Sehgal, {\it ``Neutrino Excitation of Baryon Resonances and Single Pion
Production''}, Annals of Physics {\bf 133} (1981) 79.

\bibitem{23-rs_lepton} K.M. Graczyk and J.T. Sobczyk, 
{\it ``Lepton mass effects in weak charged current single pion production''},
Phys. Rev. D{\bf 77} (2008) 053003.

\bibitem{23-rs_ff} 
K.M. Graczyk and J.T. Sobczyk,
{\it ``Form Factors in the Quark Resonance Model''}, 
Phys.Rev. D{\bf 77} (2008) 053001, Erratum-ibid. D{\bf 79} (2009) 079903.

\bibitem{23-olga_duality} O. Lalakulich, W. Melinitchouk, and E.A. Paschos, {\it ``Quark-hadron duality in 
neutrino scattering''}, Phys. Rev. C{\bf 75} (2007) 015202.

\bibitem{23-wroclaw_duality} K.M. Graczyk, C. Juszczak, and J.T. Sobczyk, {\it ``Quark-hadron duality 
in the Rein-Sehgal model''}, Nucl. Phys. A{\bf 781} (2007) 227.

\bibitem{23-duality_nucleus}	
O. Lalakulich, N. Jachowicz, Ch. Praet, and J. Ryckebusch,
{\it ``Quark-hadron duality in lepton scattering off nuclei''},
Phys. Rev. C{\bf 79} (2009) 015206.


\bibitem{B-Y} 
  A.~Bodek and U.~-k.~Yang,
  ``NUFACT09 update to the Bodek-Yang unified model for electron- and neutrino- nucleon scattering cross sections,''  AIP Conf.\ Proc.\  {\bf 1222}, 233 (2010).
  
 \bibitem{GRV98} 
  M.~Gluck, E.~Reya and A.~Vogt,
  ``Dynamical parton distributions revisited,''
  Eur.\ Phys.\ J.\ C {\bf 5}, 461 (1998) [hep-ph/9806404].
  
\bibitem{23-trvfs}
R.~Thorne and R.~Roberts, {\it ``A Practical procedure for evolving heavy flavor structure functions,''}
Phys. Lett. {\bf B421} (1998) 303. A.~D.~Martin {\em et. al.}.{\it``Estimating the effect of NNLO contributions on global parton analyses,''}
Eur. Phys. J. {\bf C18} (2000) 117.

\bibitem{23-rworld}
L.~W.~Whitlow {\em et. al},
{\it ``A Precise extraction of R = sigma-L / sigma-T from a global analysis of the SLAC deep inelastic e p and e d scattering cross-sections''},
Phys. Lett. B {\bf 250} (1990) 193.

\bibitem{23-nutev} M. Tzanov et al., {\it ``Precise measurement of neutrino and anti-neutrino differential cross sections,''}
  Phys.\ Rev.\ D {\bf 74} (2006) 012008.
 
\bibitem{23-chorus}
G. Onengut et al., \Journal{\PLB}{632}{65}{2006}

\bibitem{23-Altegoer:1997gv} 
  J.~Altegoer {\it et al.}  [NOMAD Collaboration],
  {\it ``The NOMAD experiment at the CERN SPS,''}
  Nucl.\ Instrum.\ Meth.\ A {\bf 404}, 96 (1998).

\bibitem{Petti:2006tu} 
  R.~Petti [NOMAD Collaboration],
  Nucl.\ Phys.\ Proc.\ Suppl.\  {\bf 159}, 56 (2006)
  [hep-ex/0602022].
  
\bibitem{23-Accardi:2009br} 
  A.~Accardi, M.~E.~Christy, C.~E.~Keppel, P.~Monaghan, W.~Melnitchouk, J.~G.~Morfin and J.~F.~Owens,
  {\it ``New parton distributions from large-x and low-$Q^2$ data,''}
  Phys.\ Rev.\ D {\bf 81}, 034016 (2010).

\bibitem{23-Pumplin:2002vw} 
  J.~Pumplin, D.~R.~Stump, J.~Huston, H.~L.~Lai, P.~M.~Nadolsky and W.~K.~Tung,
  {\it ``New generation of parton distributions with uncertainties from global QCD analysis,''}
  JHEP {\bf 0207}, 012 (2002).
  [hep-ph/0201195].

\bibitem{23-Stirling:2005td} 
  W.~J.~Stirling, A.~D.~Martin, R.~G.~Roberts and R.~S.~Thorne,
  {\it ``MRST parton distributions,''}
  AIP Conf.\ Proc.\  {\bf 747}, 16 (2005).

\bibitem{Tzanov:2005fu} 
  M.~Tzanov [NuTeV Collaboration],
  {\it ``NuTeV structure function measurement,''}
  AIP Conf.\ Proc.\  {\bf 792}, 241 (2005)
  [hep-ex/0507040].
  
\bibitem{23-arXiv:1012.0286} 
  K.~Kovarik, I.~Schienbein, F.~I.~Olness, J.~Y.~Yu, C.~Keppel, J.~G.~Morfin, J.~F.~Owens and T.~Stavreva,
  {\it ``Nuclear corrections in neutrino-nucleus DIS and their compatibility with global nPDF analyses,''}
Phys.\ Rev.\ Lett.\ \ {\bf 106}, 122301  (2011).

\bibitem{23-npdf}
  I.~Schienbein, J.~Y.~Yu, C.~Keppel, J.~G.~Morfin, F.~Olness and J.~F.~Owens,
  {\it ``Nuclear PDFs from neutrino deep inelastic scattering,''}
  Phys.\ Rev.\  D {\bf 77}, 054013 (2008).

\bibitem{23-owens}
  J.~F.~Owens, J.~Huston, J.~Pumplin, D.~Stump, C.~E.~Keppel, S.~Kuhlmann, J.~G.~Morfin, F.~Olness,
  ``Nuclear corrections and parton distribution functions: Lessons learned from global fitting,''
  AIP Conf.\ Proc.\  {\bf 967}, 259-263 (2007).

\bibitem{Kopeliovich:2012fu} 
  B.~Z.~Kopeliovich, J.~G.~Morfin and I.~Schmidt,
  ``Nuclear Shadowing in Electro-Weak Interactions,''
  arXiv:1208.6541 [hep-ph].

\bibitem{23-DIS12}
  \url{http://www.praktika.physik.uni-bonn.de/conferences/dis-2012?set_language=en}

\bibitem{23-Botts:1992yi}
J.~Botts et~al., {\it ``CTEQ parton distributions and flavor dependence of sea quarks,''}
  Phys.\ Lett.\ B {\bf 304}, 159 (1993).


\bibitem{23-Lai:1994bb}
H.~L. Lai et~al., {\it ``Global QCD analysis and the CTEQ parton distributions,''}
  Phys.\ Rev.\ D {\bf 51}, 4763 (1995)

  
\bibitem{23-Brodsky:2004qa}
S.~J. Brodsky, I.~Schmidt, and J.~J. Yang, {\it ``Nuclear antishadowing in neutrino deep inelastic scattering'',}
Phys. Rev., D {\bf 70}, 116003 (2004).

\bibitem{23-Qiu:2004qk} 
  J.~-W.~Qiu and I.~Vitev,
  {\it ``Nuclear shadowing in neutrino nucleus deeply inelastic scattering,''}
  Phys.\ Lett.\ B {\bf 587}, 52 (2004).

\bibitem{23-Guzey}
  V.~Guzey et al.,
  ``Impact of nuclear dependence of R=$\sigma_L/\sigma_T$ on antishadowing in nuclear structure functions,''
  arXiv:1207.0131 [hep-ph].
  
\bibitem{23-Guzey-2012}
 V.~Guzey et~al., ``Nuclear shadowing in charged-lepton-nucleus and neutrino-nucleus scattering,'' 
 work in progress.


 
\end{thebibliography}
\end{document}